\documentclass[a4paper,journal]{IEEEtran}
\usepackage{amsmath,amsfonts}
\usepackage{algorithm}
\usepackage{array}
\usepackage{subfig}
\usepackage{textcomp}
\usepackage{stfloats}
\usepackage{url}
\usepackage{verbatim}
\usepackage{graphicx}
\usepackage{amsthm}     

\theoremstyle{definition}

\theoremstyle{remark}

\newtheorem{remark}{Remark}

\usepackage{amsmath}    
\usepackage{amsfonts}   
\usepackage{bm}         
\usepackage{multirow}	
\usepackage{times} 
\usepackage{enumitem}
\usepackage{graphicx}
\usepackage{xcolor}
\usepackage{setspace}
\usepackage{hyperref}
\newcommand{\RomanNumeralCaps}[1]
{\MakeUppercase{\romannumeral #1}}

\usepackage{tikz}
\usepackage[utf8]{inputenc}
\usepackage{pgfplots}
\DeclareUnicodeCharacter{2212}{−}
\usepgfplotslibrary{groupplots,dateplot}
\usetikzlibrary{patterns,shapes.arrows}
\pgfplotsset{compat=newest}
\def\axisdefaultheight{110pt}

\begin{document}

\title{A Comparison Benchmark for Distributed Hybrid MPC Control Methods: Distributed Vehicle Platooning}
\newcommand{\mytitle}{A Comparison Benchmark for Distributed Hybrid MPC Control Methods: Distributed Vehicle Platooning}
\hypersetup{hidelinks,pdfauthor=Samuel Mallick, pdfcreator=Samuel Mallick, pdftitle=\mytitle}

\author{Samuel Mallick, Azita Dabiri, and Bart De Schutter, \IEEEmembership{Fellow, IEEE}
\thanks{This paper is part of a project that has received funding from the European Research Council (ERC) under the European Union’s Horizon
	2020 research and innovation programme (Grant agreement No. 101018826
	- CLariNet).}
\thanks{All authors are affiliated with Delft Center for Systems and Control, Delft
	University of Technology, Delft, The Netherlands (e-mail: \{s.h.mallick, a.dabiri, b.deschutter\}@tudelft.nl).}}


\IEEEpubid{0000--0000/00\$00.00~\copyright~2021 IEEE}

\maketitle

\begin{abstract}
Distributed model predictive control (MPC) is currently being investigated as a solution to the important control challenge presented by networks of hybrid dynamical systems.
However, a benchmark problem for distributed hybrid MPC is absent from the literature.
We propose distributed control of a platoon of autonomous vehicles as a comparison benchmark problem.
The problem provides a complex and adaptable case study, upon which existing and future approaches to distributed MPC for hybrid systems can be evaluated.
Two hybrid modeling frameworks are presented for the vehicle dynamics.
Five hybrid MPC controllers are then evaluated and extensively assessed on the fleet of vehicles.
Finally, we comment on the need for new efficient and high performing distributed MPC schemes for hybrid systems.

\end{abstract}

\begin{IEEEkeywords}
Autonomous vehicles, distributed model predictive control, hybrid systems, mixed integer optimization, piecewise affine systems.
\end{IEEEkeywords}

\section{Introduction}
\IEEEPARstart{D}{istributed} control of hybrid networks is an open problem in the field of systems and control \cite{lamnabhilagarrigueSystemsControlFuture2017}.
Hybrid systems are dynamical systems that combine both continuous and discrete dynamics.
Networks of hybrid systems represent many of the critical infrastructure systems in our society, such as transportation \cite{luanDecompositionDistributedOptimization2020}, energy \cite{mendesPracticalApproachHybrid2017}, and water networks \cite{vanekerenTimeInstantOptimizationHybrid2013}.
The grand societal challenge of sustainability demands the development of efficient and high-performance control approaches for these systems.
Furthermore, these control methods must be safe and reliable, as these systems are safety critical.

Model predictive control (MPC) is an optimization-based control paradigm that has been highly successful in the area of distributed control for complex systems \cite{maestreDistributedTreeBasedModel2013, suDistributedChanceConstrainedModel2019}.
MPC naturally handles multi-input-multi-output systems with constraints on the states and inputs, and is supported by a substantial body of literature on its stability and performance \cite{mayneConstrainedModelPredictive2000}.
Distributed MPC replaces a global MPC controller with local MPC controllers for each subsystem.
Distributed control is then carried out through some combination of solving the local MPC optimization problems, and communication between subsystems. 
Distributed MPC for linear systems with convex objective functions can be achieved with zero loss in performance via distributed optimization, as the global optimization problem is convex and can be solved to optimality distributively \cite{summersDistributedModelPredictive2012}.
In contrast, distributed MPC of large-scale hybrid networks is a more complex control challenge.
Compared to the linear case, the local optimization problems are highly nonlinear and non-convex, due to the combination of continuous and discrete dynamics.
Consequently, distributed optimization techniques cannot in general guarantee the reconstruction of a globally optimal control solution.

\begin{color}{black}A prominent approach to deploy MPC for several relevant classes of hybrid systems is to transform the hybrid model into mixed-logical-dynamical (MLD) form \cite{bemporadControlSystemsIntegrating1999}. 
The resulting MPC optimization problem is a mixed-integer linear/quadratic program (MILP/MIQP), for which mature solvers, e.g., Gurobi \cite{gurobi} and CPLEX \cite{cplex2009v12}, exist.
The extension to distributed MPC of hybrid systems is then to find a control solution via subsystems solving local mixed-integer programs, and cooperating by communicating the solutions between neighboring subsystems.
The local solutions can be found in parallel \cite{grossDistributedPredictiveControl2013}, iteratively solving and communicating to improve the global solution, or sequentially \cite{richardsRobustDistributedModel2007, kuwataDistributedRobustReceding2007}, where local problems are solved one after the other, with local solutions communicated down the sequence.
Another approach is to employ heuristic ways of choosing values for the integer variables while using distributed MPC to solve for values of the continuous variables \cite{mendesPracticalApproachHybrid2017, larsenModelPredictiveControl2020}.
Alternatively, in \cite{maDistributedMPCBased2023} and \cite{maRobustTubeBased2023} distributed MPC approaches for piecewise affine (PWA) systems, a class of hybrid systems, are proposed where the couplings between subsystems are treated as disturbances.
The approaches are limited to regulation problems in which a robust terminal set exists.
 \IEEEpubidadjcol
 
An alternative set of approaches are based on distributed mixed-integer optimization, which enables distributed hybrid MPC by solving a global mixed-integer hybrid MPC problem distributively among subsystems.
In \cite{vujanicDecompositionMethodLarge2016} a distributed solution for large-scale MILPs is proposed using a dual formulation of the problem, guaranteeing a primally feasible solution and bounded suboptimality.  
This idea was then extended in \cite{falsoneDecentralizedApproachMultiAgent2019} to include finite-time feasibility.
In \cite{camisaDistributedPrimalDecomposition2022} a distributed large-scale MILP approach is proposed based on primal decomposition.
These works, however, consider only inequality constraint coupling, and assume that the number of subsystems far exceeds the number of coupling constraints.
These approaches are hence unsuitable for control problems with coupling in the cost or dynamics as, while cost and dynamics coupling can be converted to constraint coupling via introducing auxiliary variables \cite{summersDistributedModelPredictive2012}, the number of constraints then far exceeds the number of subsystems.
Finally, while the alternating direction method of multipliers (ADMM) is only guaranteed to converge for convex problems \cite{boydDistributedOptimizationStatistical2010}, it has been used as a distributed solution approach for mixed-integer programs.
In \cite{liuDistributedSolutionMixedInteger2022} a modified ADMM procedure is presented for mixed-integer problems; however, the coupling between subproblems is restricted to only the binary variables.
Additionally, in the context of distributed hybrid MPC, ADMM has been used as a heuristic method for solving the global optimization problem distributively without guarantees \cite{luanDecompositionDistributedOptimization2020, beltranUnitCommitmentAugmented2002}.
\end{color}

The methods referenced above are demonstrated on a range of different case studies, with the nature and complexity of the control problem varying from example to example.
\begin{color}{black}To the best of the authors' knowledge, a standardized benchmark problem, based on a relevant real-world challenge, for evaluating the effectiveness of distributed hybrid MPC approaches is missing from the literature.\end{color}
The goal of this paper is to propose such a benchmark. 

We introduce the platooning of a fleet of vehicles, where vehicle gear shifting are explicitly optimized, as the benchmark problem.
Control of a single SMART\footnote{SMART is a German vehicle manufacturer \cite{SMART}.} vehicle has been proposed as a comparative benchmark problem for centralized MPC of hybrid systems \cite{coronaAdaptiveCruiseControl2008}.
Vehicle platooning in the context of distributed control has been explored using non-hybrid models in \cite{liDynamicalModelingDistributed2017, zhengDistributedModelPredictive2017, liuDistributedModelPredictive2019}.
In these works the inherent hybrid nature of discrete gear choices is either assumed to be absent or is neglected.
\begin{color}{black}While vehicles with continuous-variable-transmission (CVT) exist, where there is no notion of a discrete gear, these account for a very small share of vehicles only \cite{mobilityforesights2024cvt}.
Moreover, explicit gear management has been identified as important for fuel consumption, effecting the efficiency of the engine, and for tracking and platooning behavior, as gear shifts can introduce large deviations in speed and inter-vehicular distance \cite{turriGearManagementFuelEfficient2016}.\end{color}
For control of a single autonomous vehicle, optimization of both continuous control inputs and discrete gear choices has been addressed in \cite{liEcologicalAdaptiveCruise2020, shaoVehicleSpeedGear2021}.
However, these approaches do not extend to platoons of vehicles, and they are inherently centralized control approaches.
The only previous work to consider control of gear shifts in platooning is  \cite{turriGearManagementFuelEfficient2016}, where the gear shifting control is decoupled from the longitudinal vehicle control. 
Finally, in \cite{zhangHybridMPCSystem2022} a hybrid distributed MPC approach is formulated for platoon control; however, the hybrid nature of the problem comes from lane changes, while the vehicles have non-hybrid dynamics.

\begin{color}{black}The current paper makes the following contributions to the state of the art:
\begin{itemize}
	\item A platooning problem is proposed as a useful comparison benchmark problem for distributed hybrid MPC approaches. 
	The problem scrutinizes, for distributed hybrid MPC strategies, the handling of many discrete decision variables, through the (predicted) gear transitions, and the effective coordination and communication between distributed controllers, with heavy coupling between controllers arising from position tracking and coupled safety constraints.
	\item Two hybrid models are introduced for the vehicles with explicit gear management.
	The platooning problem for hybrid vehicles with gear transitions is formulated, with a variety of \textit{tuning knobs} determining the complexity of the control challenge.
	\item An open-source code base for the benchmark problem is provided, where alternative models, controllers, and tasks can be evaluated. 
	\item As a first use of the benchmark, five representative existing control strategies are extensively evaluated and their performance, complexity, and characteristics are discussed.
\end{itemize}
\end{color}

This paper is organized as follows.
Section \ref{sec:model} introduces two hybrid models for vehicles with explicit gear management.
Section \ref{sec:benchmark} formulates the benchmark control problem.
Section \ref{sec:controllers} gives a brief overview of each of the control methods to be evaluated.
Section \ref{sec:experiments} provides simulation results, and compares and assesses the behaviors of the controllers.
Finally, in Section \ref{sec:conclussion} we provide concluding remarks and discuss the open challenges highlighted by the benchmark case study.

\section{Vehicle Models} \label{sec:model}
In this section we detail two modeling approaches for the dynamics of a vehicle with explicit gear management.
\begin{color}{black}These models directly use the friction model from \cite{coronaAdaptiveCruiseControl2008}; however, the presented gear models are novel.  \end{color}
Whereas \cite{coronaAdaptiveCruiseControl2008} introduces errors into the gear traction values, in order to express the traction as an affine function of the gear, in this work we present two models that use the true traction values.

\subsection{Models}
Considering a vehicle driving forwards, an accurate model of its dynamics is 
\begin{equation}
	\label{eq:ODE}
	m\ddot{s}(t) + c \dot{s}^2 + \mu m g = b(j, \dot{s})u(t),
\end{equation}
where $s(t)$ is the position at time $t$, $j \in \{1, \dots, 6\}$ is the selected gear, and $b(j, \dot{s})u(t)$ is a traction force that is proportional to the normalized throttle position $u(t)$.
In \eqref{eq:ODE} $g$ is gravitational acceleration, $\mu$ is the Coulomb friction coefficient, $c$ is the viscous friction coefficient, and $m$ is the vehicle mass (values in Table \ref{tab:constraints}).
Defining the state as position and velocity via $x = \begin{bmatrix}
	s & \dot{s}
\end{bmatrix}^\top$, the dynamics are expressed as
\begin{equation}
	\dot{x} = A(x) + B(j, x)u
\end{equation}
where 
\begin{equation}
	A(x) = \begin{bmatrix}
		x_2 \\ -(1/m)f(x_2) - \mu g
	\end{bmatrix}, \: B(j, x) = \begin{bmatrix}
	0 \\ b(j, x_2)/m
	\end{bmatrix},
\end{equation}
and with $f(x_2) = c x_2^2$.
The dynamics are nonlinear, due to the quadratic friction, and hybrid, due to the discrete variable $j$.
The quadratic friction $f(x_2)$ is depicted in 
Figure \ref{fig:fric} and the traction force $b(j, x_2)$ in Figure \ref{fig:gears_full}.
Table \ref{tab:gears} gives the maximum traction value for each gear and the velocity ranges for which this maximum traction is constant.

\begin{table}
	\centering
	\caption{Gear traction}
	\label{tab:gears}
	\begin{tabular}{cccc}
		\hline
		\textbf{Gear} $j$ & $\substack{\textbf{Traction force} \\ b(j) (N)}$ & $\substack{\textbf{Min. vel.} \\ (m/s)}$ & $\substack{\textbf{Max. vel.} \\ (m/s)}$ \\
		\hline
		\RomanNumeralCaps{1} & 4057 &  3.94 & 9.46\\
		\RomanNumeralCaps{2} & 2945 & 5.43 & 13.04\\
		\RomanNumeralCaps{3} & 2116 & 7.56 & 18.15\\
		\RomanNumeralCaps{4} & 1607 & 9.96 & 23.90\\
		\RomanNumeralCaps{5} & 1166 & 13.70 & 32.93\\
		\RomanNumeralCaps{6} & 838 & 19.10 & 45.84\\
		\hline
	\end{tabular}
\end{table}
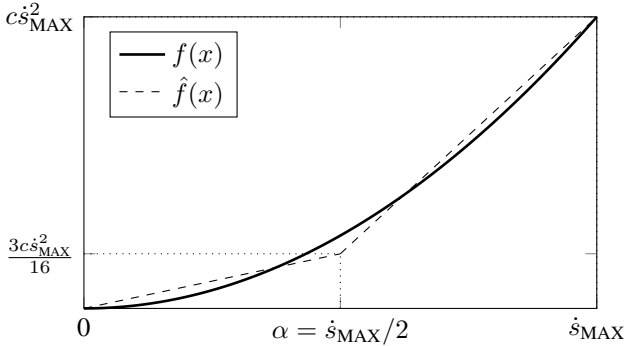
\begin{figure}
	\centering
%
%
\begin{tikzpicture}

\begin{axis}[%
legend cell align={left},
legend style={
	fill opacity=0.8,
	draw opacity=1,
	text opacity=1,
	at={(0.05, 0.95)},
	anchor=north west,
},
width=0.8*\axisdefaultwidth,
height=\axisdefaultheight,
at={(0\textwidth,0\textwidth)},
scale only axis,
clip=false,
xmin=0,
xmax=50,
xtick={0, 22.92, 45.84},
xmax=45.84,
xticklabels={0, \(\alpha=\Dot{s}_\text{MAX}/2\), \(\Dot{s}_\text{MAX}\)},
ymin=0,
ymax=1200,
ytick={196.9974, 1050.7},
ymax=1051,
yticklabels={\(\frac{3c\Dot{s}^2_\text{MAX}}{16}\), \(c\Dot{s}^2_\text{MAX}\)},
axis background/.style={fill=white}
]
\addplot [color=black, line width=1.0pt]
  table[row sep=crcr]{%
0	0\\
0.463030303030303	0.107198530762167\\
0.926060606060606	0.428794123048669\\
1.38909090909091	0.964786776859504\\
1.85212121212121	1.71517649219467\\
2.31515151515152	2.67996326905418\\
2.77818181818182	3.85914710743802\\
3.24121212121212	5.25272800734619\\
3.70424242424242	6.8607059687787\\
4.16727272727273	8.68308099173554\\
4.63030303030303	10.7198530762167\\
5.09333333333333	12.9710222222222\\
5.55636363636364	15.4365884297521\\
6.01939393939394	18.1165516988063\\
6.48242424242424	21.0109120293848\\
6.94545454545455	24.1196694214876\\
7.40848484848485	27.4428238751148\\
7.87151515151515	30.9803753902663\\
8.33454545454546	34.7323239669422\\
8.79757575757576	38.6986696051423\\
9.26060606060606	42.8794123048669\\
9.72363636363636	47.2745520661157\\
10.1866666666667	51.8840888888889\\
10.649696969697	56.7080227731864\\
11.1127272727273	61.7463537190083\\
11.5757575757576	66.9990817263545\\
12.0387878787879	72.466206795225\\
12.5018181818182	78.1477289256198\\
12.9648484848485	84.043648117539\\
13.4278787878788	90.1539643709826\\
13.8909090909091	96.4786776859504\\
14.3539393939394	103.017788062443\\
14.8169696969697	109.771295500459\\
15.28	116.7392\\
15.7430303030303	123.921501561065\\
16.2060606060606	131.318200183655\\
16.6690909090909	138.929295867769\\
17.1321212121212	146.754788613407\\
17.5951515151515	154.794678420569\\
18.0581818181818	163.048965289256\\
18.5212121212121	171.517649219467\\
18.9842424242424	180.200730211203\\
19.4472727272727	189.098208264463\\
19.910303030303	198.210083379247\\
20.3733333333333	207.536355555556\\
20.8363636363636	217.077024793388\\
21.2993939393939	226.832091092746\\
21.7624242424242	236.801554453627\\
22.2254545454545	246.985414876033\\
22.6884848484849	257.383672359963\\
23.1515151515152	267.996326905418\\
23.6145454545455	278.823378512397\\
24.0775757575758	289.8648271809\\
24.5406060606061	301.120672910927\\
25.0036363636364	312.590915702479\\
25.4666666666667	324.275555555556\\
25.929696969697	336.174592470156\\
26.3927272727273	348.288026446281\\
26.8557575757576	360.61585748393\\
27.3187878787879	373.158085583104\\
27.7818181818182	385.914710743802\\
28.2448484848485	398.885732966024\\
28.7078787878788	412.071152249771\\
29.1709090909091	425.470968595041\\
29.6339393939394	439.085182001837\\
30.0969696969697	452.913792470156\\
30.56	466.9568\\
31.0230303030303	481.214204591368\\
31.4860606060606	495.686006244261\\
31.9490909090909	510.372204958678\\
32.4121212121212	525.272800734619\\
32.8751515151515	540.387793572084\\
33.3381818181818	555.717183471075\\
33.8012121212121	571.260970431589\\
34.2642424242424	587.019154453627\\
34.7272727272727	602.99173553719\\
35.190303030303	619.178713682277\\
35.6533333333333	635.580088888889\\
36.1163636363636	652.195861157025\\
36.5793939393939	669.026030486685\\
37.0424242424242	686.07059687787\\
37.5054545454545	703.329560330579\\
37.9684848484848	720.802920844812\\
38.4315151515152	738.49067842057\\
38.8945454545455	756.392833057851\\
39.3575757575758	774.509384756658\\
39.8206060606061	792.840333516988\\
40.2836363636364	811.385679338843\\
40.7466666666667	830.145422222222\\
41.209696969697	849.119562167126\\
41.6727272727273	868.308099173554\\
42.1357575757576	887.711033241506\\
42.5987878787879	907.328364370983\\
43.0618181818182	927.160092561983\\
43.5248484848485	947.206217814509\\
43.9878787878788	967.466740128559\\
44.4509090909091	987.941659504132\\
44.9139393939394	1008.63097594123\\
45.3769696969697	1029.53468943985\\
45.84	1050.6528\\
};
\addlegendentry{\(f(x)\)}
\addplot [color=black, dashed]
  table[row sep=crcr]{%
0	0\\
22.92	196.9974\\
};
\addlegendentry{\(\Hat{f}(x)\)}
\addplot [color=black, dashed, forget plot]
  table[row sep=crcr]{%
22.92	196.9974\\
45.84	1050.6528\\
};
\addplot [color=black, dotted, forget plot]
  table[row sep=crcr]{%
45.84	0\\
45.84	1050.6528\\
};
\addplot [color=black, dotted, forget plot]
  table[row sep=crcr]{%
0	1050.6528\\
45.84	1050.6528\\
};
\addplot [color=black, dotted, forget plot]
  table[row sep=crcr]{%
22.92	0\\
22.92	196.9974\\
};
\addplot [color=black, dotted, forget plot]
  table[row sep=crcr]{%
0	196.9974\\
22.92	196.9974\\
};
\end{axis}
\end{tikzpicture}%
	\caption{PWA friction approximation. True quadratic function (solid), piecewise approximation (dashed).}
	\label{fig:fric}
\end{figure}

A common approach to `hybridize' a nonlinearity is to use a PWA approximation, replacing a nonlinear curve with a series of affine pieces.
As in \cite{coronaAdaptiveCruiseControl2008}, we approximate $f$ with the PWA function $\hat{f}$ in Figure \ref{fig:fric}, using two affine regions,
\begin{equation}
	\label{eq:pwa_fric}
	\hat{f}(x_2) = \begin{cases}
		a_1 x_2 + c_1 & x_2 \leq \alpha \\
		a_2 x_2 + c_2 & x_2 > \alpha
	\end{cases},
\end{equation}
where $\alpha = \dot{s}_{\text{MAX}}/2$, and $ \dot{s}_{\text{MAX}}$ is a maximum velocity, defined in Table \ref{tab:constraints}.
A PWA approximation of $A(x)$ is then
\begin{equation}
	A_\text{PWA}(x) = \begin{bmatrix}
		x_2 \\ -\hat{f}(x_2)/m - \mu g
	\end{bmatrix}.
\end{equation}

We explore two models for the hybrid gear dynamics $B(j, x)$.
The first approach considers the gear choice as dependent on the velocity, giving a PWA model of $B(j, x)$.
The second approach will consider the gear choice as an independent discrete decision variable.

\subsubsection{PWA gear model}
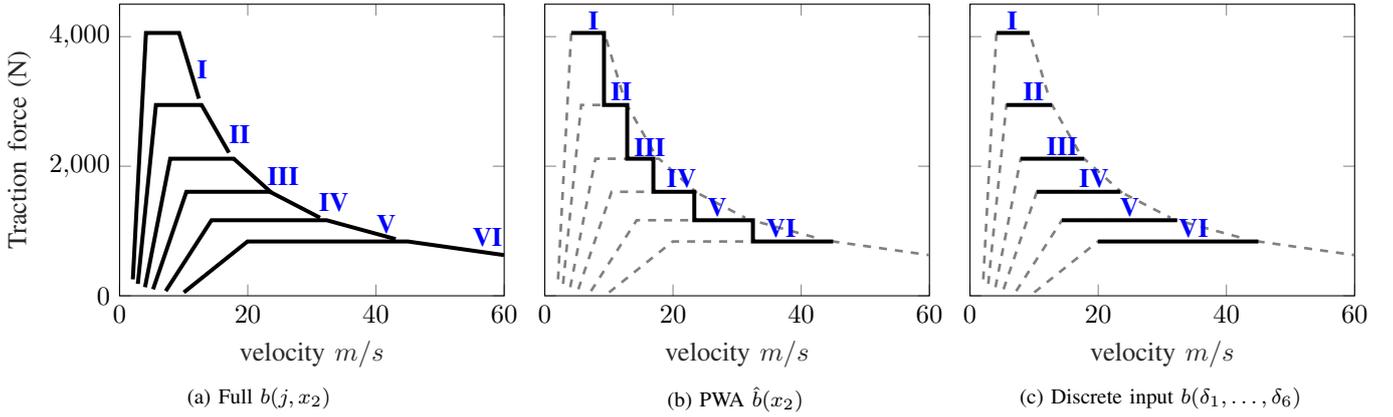
\begin{figure*}
	\centering
	\subfloat[Full $b(j, x_2)$ \label{fig:gears_full}]{
%
%
\begin{tikzpicture}
	
\newcommand{\lw}{1.5pt}

\begin{axis}[%
width=0.6*\axisdefaultwidth,
height=\axisdefaultheight,
at={(0\textwidth,0\textwidth)},
scale only axis,
xmin=0,
xmax=60,
xlabel style={font=\color{white!15!black}},
xlabel={velocity \(m/s\)},
ymin=0,
ymax=4500,
ylabel style={font=\color{white!15!black}},
ylabel={Traction force (N)},
axis background/.style={fill=white}
]
\addplot [color=black, line width=\lw, forget plot]
  table[row sep=crcr]{%
2.0706	253.54\\
4.12158	4056.7\\
9.29	4056.7\\
12.38	3042.52\\
};
\addplot [color=black, line width=\lw, forget plot]
  table[row sep=crcr]{%
2.85	184\\
5.675	2944.75\\
12.7956	2944.75\\
17.06	2208.55\\
};
\addplot [color=black, line width=\lw, forget plot]
  table[row sep=crcr]{%
3.9705	132.22\\
7.90316	2115.6\\
17.8105	2115.6\\
23.7474	1586.7\\
};
\addplot [color=black, line width=\lw, forget plot]
  table[row sep=crcr]{%
5.228	100.415\\
10.42	1605\\
23.454	1605\\
31.2704	1205\\
};
\addplot [color=black, line width=\lw, forget plot]
  table[row sep=crcr]{%
7.203	72.88\\
14.335	1166\\
32.31	1166\\
43.0802	874.7\\
};
\addplot [color=black, line width=\lw, forget plot]
  table[row sep=crcr]{%
10.027	52.4\\
19.956	838\\
44.978	838\\
59.9715	628.3\\
};
\node[right, align=left, inner sep=0, font=\bfseries\color{blue}]
at (axis cs:12,3500) {I};
\node[right, align=left, inner sep=0, font=\bfseries\color{blue}]
at (axis cs:17,2500) {II};
\node[right, align=left, inner sep=0, font=\bfseries\color{blue}]
at (axis cs:23,1850) {III};
\node[right, align=left, inner sep=0, font=\bfseries\color{blue}]
at (axis cs:31,1450) {IV};
\node[right, align=left, inner sep=0, font=\bfseries\color{blue}]
at (axis cs:40,1100) {V};
\node[right, align=left, inner sep=0, font=\bfseries\color{blue}]
at (axis cs:55,900) {VI};
\end{axis}
\end{tikzpicture}%
}
	\subfloat[PWA $\hat{b}(x_2)$ \label{fig:gears_pwa}]{
%
%
\begin{tikzpicture}
	
	\newcommand{\lw}{1.5pt}

\begin{axis}[%
width=0.6*\axisdefaultwidth,
height=\axisdefaultheight,
at={(0\textwidth,0\textwidth)},
scale only axis,
xmin=0,
xmax=60,
xlabel style={font=\color{white!15!black}},
xlabel={velocity \(m/s\)},
ymin=0,
ymax=4500,
ylabel style={font=\color{white!15!black}},
yticklabels={},
axis background/.style={fill=white}
]
\addplot [color=gray, dashed, line width=1.0pt, forget plot]
  table[row sep=crcr]{%
2.0706	253.54\\
4.12158	4056.7\\
9.29	4056.7\\
12.38	3042.52\\
};
\addplot [color=gray, dashed, line width=1.0pt, forget plot]
  table[row sep=crcr]{%
2.85	184\\
5.675	2944.75\\
12.7956	2944.75\\
17.06	2208.55\\
};
\addplot [color=gray, dashed, line width=1.0pt, forget plot]
  table[row sep=crcr]{%
3.9705	132.22\\
7.90316	2115.6\\
17.8105	2115.6\\
23.7474	1586.7\\
};
\addplot [color=gray, dashed, line width=1.0pt, forget plot]
  table[row sep=crcr]{%
5.228	100.415\\
10.42	1605\\
23.454	1605\\
31.2704	1205\\
};
\addplot [color=gray, dashed, line width=1.0pt, forget plot]
  table[row sep=crcr]{%
7.203	72.88\\
14.335	1166\\
32.31	1166\\
43.0802	874.7\\
};
\addplot [color=gray, dashed, line width=1.0pt, forget plot]
  table[row sep=crcr]{%
10.027	52.4\\
19.956	838\\
44.978	838\\
59.9715	628.3\\
};
\addplot [color=black, line width=\lw, forget plot]
  table[row sep=crcr]{%
4.12158	4056.7\\
9.2353	4056.7\\
9.2353	2944.75\\
12.85683	2944.75\\
12.85683	2115.6\\
16.937	2115.6\\
16.937	1605\\
23.3225	1605\\
23.3225	1166\\
32.467	1166\\
32.467	838\\
44.978	838\\
};
\node[right, align=left, inner sep=0, font=\bfseries\color{blue}]
at (axis cs:6.706,4250.7) {I};
\node[right, align=left, inner sep=0, font=\bfseries\color{blue}]
at (axis cs:10.235,3150.75) {II};
\node[right, align=left, inner sep=0, font=\bfseries\color{blue}]
at (axis cs:13.857,2300.6) {III};
\node[right, align=left, inner sep=0, font=\bfseries\color{blue}]
at (axis cs:18.937,1805) {IV};
\node[right, align=left, inner sep=0, font=\bfseries\color{blue}]
at (axis cs:25.322,1366) {V};
\node[right, align=left, inner sep=0, font=\bfseries\color{blue}]
at (axis cs:34.467,1040) {VI};
\end{axis}
\end{tikzpicture}%
}
	\subfloat[Discrete input $b(\delta_1, \dots, \delta_6)$ \label{fig:gears_mld}]{
%
%
\begin{tikzpicture}
	
\newcommand{\lw}{1.5pt}

\begin{axis}[%
width=0.6*\axisdefaultwidth,
height=\axisdefaultheight,
at={(0\textwidth,0\textwidth)},
scale only axis,
xmin=0,
xmax=60,
xlabel style={font=\color{white!15!black}},
xlabel={velocity \(m/s\)},
ymin=0,
ymax=4500,
ylabel style={font=\color{white!15!black}},
yticklabels={},
axis background/.style={fill=white}
]
\addplot [color=gray, dashed, line width=1.0pt, forget plot]
  table[row sep=crcr]{%
2.0706	253.54\\
4.12158	4056.7\\
9.29	4056.7\\
12.38	3042.52\\
};
\addplot [color=gray, dashed, line width=1.0pt, forget plot]
  table[row sep=crcr]{%
2.85	184\\
5.675	2944.75\\
12.7956	2944.75\\
17.06	2208.55\\
};
\addplot [color=gray, dashed, line width=1.0pt, forget plot]
  table[row sep=crcr]{%
3.9705	132.22\\
7.90316	2115.6\\
17.8105	2115.6\\
23.7474	1586.7\\
};
\addplot [color=gray, dashed, line width=1.0pt, forget plot]
  table[row sep=crcr]{%
5.228	100.415\\
10.42	1605\\
23.454	1605\\
31.2704	1205\\
};
\addplot [color=gray, dashed, line width=1.0pt, forget plot]
  table[row sep=crcr]{%
7.203	72.88\\
14.335	1166\\
32.31	1166\\
43.0802	874.7\\
};
\addplot [color=gray, dashed, line width=1.0pt, forget plot]
  table[row sep=crcr]{%
10.027	52.4\\
19.956	838\\
44.978	838\\
59.9715	628.3\\
};
\addplot [color=black, line width=\lw, forget plot]
  table[row sep=crcr]{%
4.12158	4056.7\\
9.29	4056.7\\
};
\addplot [color=black, line width=\lw, forget plot]
  table[row sep=crcr]{%
5.675	2944.75\\
12.7956	2944.75\\
};
\addplot [color=black, line width=\lw, forget plot]
  table[row sep=crcr]{%
7.90316	2115.6\\
17.8105	2115.6\\
};
\addplot [color=black, line width=\lw, forget plot]
  table[row sep=crcr]{%
10.42	1605\\
23.454	1605\\
};
\addplot [color=black, line width=\lw, forget plot]
  table[row sep=crcr]{%
14.335	1166\\
32.31	1166\\
};
\addplot [color=black, line width=\lw, forget plot]
  table[row sep=crcr]{%
19.956	838\\
44.978	838\\
};
\node[right, align=left, inner sep=0, font=\bfseries\color{blue}]
at (axis cs:5.706,4250.7) {I};
\node[right, align=left, inner sep=0, font=\bfseries\color{blue}]
at (axis cs:8.235,3150.75) {II};
\node[right, align=left, inner sep=0, font=\bfseries\color{blue}]
at (axis cs:11.857,2315.6) {III};
\node[right, align=left, inner sep=0, font=\bfseries\color{blue}]
at (axis cs:16.937,1805) {IV};
\node[right, align=left, inner sep=0, font=\bfseries\color{blue}]
at (axis cs:23.322,1366) {V};
\node[right, align=left, inner sep=0, font=\bfseries\color{blue}]
at (axis cs:32.467,1040) {VI};
\end{axis}
\end{tikzpicture}%
}
	\caption{Gear models.}
\end{figure*}

For a PWA model of the gear dynamics, we consider the regions of the traction curves in Figure \ref{fig:gears_full} where the traction is constant.
The velocity is then partitioned, and a gear is chosen for each region, such that there is a one-to-one mapping from velocity to gear.
We take the mid-point of a gear's constant velocity range as the lower bound for the PWA region associated with that gear.
The PWA gear model, depicted in Figure \ref{fig:gears_pwa}, is then $B_{\text{PWA}}(x) = \begin{bmatrix}
	0 & \hat{b}(x_2)/m
\end{bmatrix}^\top$ where
\begin{equation}
	 \hat{b}(x_2) = \begin{cases}
		b_{1,\text{H}} & v_{1, \text{L}} \leq x_2 < \frac{v_{2, \text{L}}+v_{2, \text{H}}}{2} \\
		b_{2,\text{H}} & \frac{v_{2, \text{L}}+v_{2, \text{H}}}{2} \leq x_2 < \frac{v_{3, \text{L}}+v_{3, \text{H}}}{2} \\
		\vdots \\
		b_{6,\text{H}} & \frac{v_{6, \text{L}}+v_{6, \text{H}}}{2} \leq x_2 < v_{6, \text{H}}
	\end{cases},
\end{equation}
and $b_{j,\text{H}}$, $v_{j, \text{L}}$, and $ v_{j, \text{H}}$, are the maximum traction, minimum velocity, and maximum velocity bounds for gear $j$, as given in Table \ref{tab:gears}.
The PWA gear model no longer includes a discrete decision variable $j$, and is a function only of the state $x$.
Combining $B_{\text{PWA}}(x)$ with $A_\text{PWA}(x)$ gives the PWA model
\begin{equation}
	\label{eq:PWA_mod}
	\dot{x} = A_\text{PWA}(x) + B_{\text{PWA}}(x) u
\end{equation}
with seven affine regions determined by the velocity (six from the gears and an extra from the friction).
We will refer to this model as Model \RomanNumeralCaps{1}.
This model can serve as the prediction model in an MPC controller for the vehicle, and can be converted into a variety of equivalent hybrid models \cite{heemelsEquivalenceHybridDynamical2001a}, resulting in different MPC optimization problems, e.g., an MLD model, giving a mixed-integer MPC problem \cite{bemporadControlSystemsIntegrating1999}.

\subsubsection{Discrete-input gear model}
The second gear model we consider describes the gear choice as a discrete input.
Again the traction curves are restricted to the regions of constant traction, and each gear is restricted to operate only in these regions.
However, unlike the PWA approximation, the mapping from velocity to gear is not one-to-one.
To represent the gear choice, six binary variables $\delta_1 - \delta_6$ are introduced, one for each gear.
When $\delta_j$ is equal to one, gear $j$ is chosen, with the constraint
\begin{equation}
	\sum_{j=1}^6 \delta_j = 1
\end{equation}
ensuring that only one gear is active.
Figure \ref{fig:gears_mld} depicts the discrete-input gear model
\begin{equation}
	b(\delta_1, \dots, \delta_6) = \sum_{j=1}^6 \delta_j b_{j,\text{H}},
\end{equation}
with the $B$ matrix is then
\begin{equation}
	B_{\text{DISC}}(\delta_1, \dots, \delta_6) = \begin{bmatrix}
		0 \\ b(\delta_1, \dots, \delta_6)/m
		\end{bmatrix}.
\end{equation}
The nonlinearity originating from the multiplication of binary and continuous decision variables in $B_\text{DISC}(\delta_1, \dots, \delta_6) u$ can be reformulated as the linear expression $B_\text{MLD}(\delta_1, \dots, \delta_6, u)$ through the addition of auxiliary variables and mixed-integer linear constraints (see \cite{bemporadControlSystemsIntegrating1999} for details).
Finally, restricting the gears to operate within the velocity regions of constant traction is achieved via the inclusion of additional mixed-integer constraints that encode the logical expressions, e.g.,
\begin{equation}
	\big(\delta_j = 1 \implies x_2 \leq v_{j, \text{H}}\big) \iff x_2 - v_{j, \text{H}} \leq M_\text{H}(1-\delta_j),
\end{equation}
where $M_\text{H} = \max_x (x_2 - v_{1, \text{H}})$, and
\begin{equation}
	\big(\delta_j = 1 \implies x_2 \geq v_{j, \text{L}}\big) \iff v_{j, \text{L}} - x_2 \leq M_\text{L}(1-\delta_j),
\end{equation}
where $M_\text{L} = \max_x (v_{1, \text{L}} - x_2)$.

Converting $A_\text{PWA}(x)$ to MLD form (again, see \cite{bemporadControlSystemsIntegrating1999} for details) $A_\text{MLD}(x)$ and combining with $B_{\text{MLD}}(\delta_1, \dots, \delta_6, u)$ gives the MLD model
\begin{equation}
\label{eq:MLD_mod}
\begin{aligned}
	\dot{x} &= A_\text{MLD}(x) + B_{\text{MLD}}(\delta_1, \dots, \delta_6, u) \\
	&\text{s.t.} \quad \text{mixed-integer equations},
\end{aligned}
\end{equation}
with eight binary variables (six from the gears and two from the PWA regions of the friction).
We will refer to this model as Model \RomanNumeralCaps{2}.
This model can serve as the prediction model in an MPC controller, resulting in a mixed-integer optimization problem.
However, as the velocity to gear mapping is not one-to-one, the model cannot be converted into PWA form.

In an MPC context a discrete-time model is required.
Both models are discretized using the forward Euler method and a sampling time $T$.
In the following, all references to models refer to the discrete-time versions, and $k$ represents the discrete-time step counter.

\section{The Benchmark Problem} \label{sec:benchmark}
\begin{color}{black}
Consider the control problem where a platoon of $M$ vehicles track each other's position and velocity.
Define the set of vehicles as $\mathcal{M} = \{1, \dots, M\}$, with vehicle $i$ having mass $m_i$, and state $x^{(i)}$ containing its position and velocity.
Without loss of generality we assume the set is sorted by the position of the vehicles, i.e., for the first vehicle $i = 1$, and vehicle $i$ is behind vehicle $i - 1$.
Only longitudinal motion is considered, i.e., there is only one lane with no overtaking.
One vehicle $l \in \mathcal{M}$, henceforth referred to as the leader, is provided with a reference trajectory $r = \{r(k)\}_{k \geq 0}$, where $r(k) \in \mathbb{R}^2$ is the desired state of the leader vehicle at time step $k$.
All other vehicles track the states of the preceding and succeeding vehicles, with the desired difference between the state of vehicle $i$ and vehicle $i-1$ given by an inter-vehicle spacing policy $\eta(x^{(i)}) \in \mathbb{R}^2$.
It is assumed that each vehicle can take error-free measurements of the current position and velocity of the preceding and succeeding vehicles.

Within the scope of this problem there a set of \textit{tuning knobs} that alter the complexity of the control challenge:
\begin{itemize}
	\item The first is the number of vehicles $M$, with more vehicles presenting a larger global system, and more subsystems that must coordinate.
	\item The second is the reference trajectory $r$, which determines the overall behavior of the platoon, and can trigger smooth or aggressive behaviors.
	\item The third is the inter-vehicle spacing policy $\eta$. 
	Constant-distance, $\eta(x^{(i)}) = \begin{bmatrix}d_0& 0\end{bmatrix}^\top$, and velocity-dependent, $\eta(x^{(i)}) = \begin{bmatrix}t_0 x_2^{(i)} + d_0 & 0\end{bmatrix}^\top$, spacing policies are considered, where $d_0$ and $t_0$ are a fixed distance and time, respectively.
	A velocity-dependent spacing policy increases the effective coupling between vehicles, as vehicle $i-1$ maintains a spacing from vehicle $i$ that depends on the velocity of vehicle $i$.
	\item The fourth is vehicle inhomogeneity, introduced through variations in the masses $m_i$.
	Vehicle inhomogeneity enhances the need for cooperation, as lighter vehicles can accelerate faster, and without coordination will not be trackable by heavier vehicles.
	\item The fifth is leader position. 
	Platooning typically considers the front vehicle as the leader, simplifying the problem, as the front vehicle is not restricted by a preceding vehicle.
	By assigning the leader role to a vehicle within the platoon, $l \neq 1$, the reference tracking is more challenging.
\end{itemize} \end{color}
\begin{remark}
	We note that varying communication topologies have been explored in \cite{liDynamicalModelingDistributed2017}.
	\begin{color}{black}Additionally, predictive control under state measurement errors has been considered in \cite{lanDataDrivenRobustPredictive2023}.
	In building this benchmark case study, we allow the communication topology to vary with the controllers, with the level and type of communication being a point of comparison.
	We also assume no measurement errors, to maintain the focus on nominal controller performance.\end{color}
\end{remark}
\begin{remark}
	A common consideration in platooning is stability for each vehicle \cite{zhengDistributedModelPredictive2017, dunbarDistributedRecedingHorizon2012}, and the string stability of the network \cite{liDynamicalModelingDistributed2017}.
	In the current paper, the focus is on defining a benchmark for the performance of distributed hybrid MPC controllers.
	As such we neglect these components, focusing on the tracking performance and the communication and computation time requirements of the controllers.
	Addressing the remaining components in the context of this benchmark problem will be the focus of future work.
\end{remark}

\begin{figure*}
	\centering
	\includegraphics[width=0.8\textwidth]{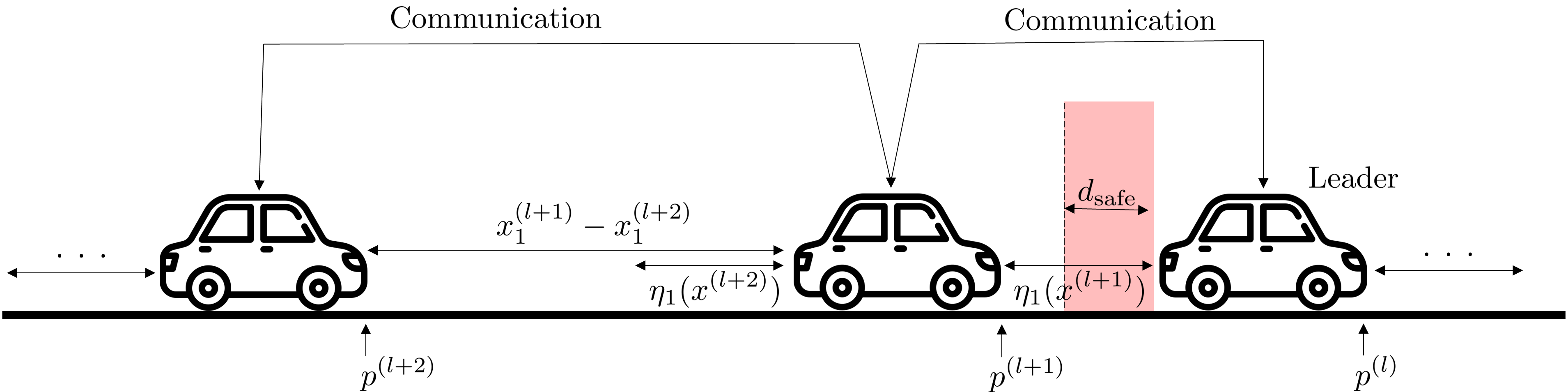}
	\caption{Platoon control.}
	\label{fig:cars}
\end{figure*}

\subsection{Constraints} \label{sec:constraints}
The states and inputs of the vehicles are subject to constraints.
Numerical values for the constraint coefficients used in our experiments are given in Table \ref{tab:constraints}.
Velocity and acceleration constraints are imposed for safety and comfort:
\begin{subequations} \label{eq:vel_constraints}
	\begin{align}
		&v_{\text{MIN}} \leq x_2^{(i)}(k) \leq v_{\text{MAX}} \label{eq:vel_c} \\
		&a_{\text{MIN}}T \leq x_2^{(i)}(k+1) - x_2^{(i)}(k) \leq a_{\text{MAX}}T, \label{eq:acc_c}
	\end{align}
\end{subequations}
where $v_{\text{MIN}} > 0$, $v_{\text{MAX}} > 0$, $a_{\text{MIN}} < 0$, and $a_{\text{MAX}} > 0$ are velocity and acceleration bounds respectively.
Note that the lower bound on velocity is to maintain the validity of the constant traction approximations in the models, and to ensure the vehicles drive forwards.
The normalized throttle input must be constrained as 
\begin{equation} \label{eq:throttle_constraints}
	-u_{\text{MAX}} \leq u^{(i)}(k) \leq u_{\text{MAX}}.
\end{equation}
The vehicles (excluding the front vehicle) must maintain a safe distance $d_{\text{safe}}$ from the preceding vehicle:
\begin{equation}\label{eq:safe_constraints}
	x_1^{(i)}(k) \leq x_1^{(i - 1)}(k) - d_{\text{safe}}.
\end{equation}
Finally, we impose non-restrictive constraints on the position, as the PWA to MLD model conversion requires that the states are bounded \cite{bemporadControlSystemsIntegrating1999}:
\begin{equation} \label{eq:pos_constraints}
	p_{\text{MIN}} \leq x_1^{(i)}(k) \leq p_{\text{MAX}},
\end{equation}
where  $p_{\text{MIN}}$ and $p_{\text{MAX}}$ are position bounds.
This is not restrictive as, in an MPC approach, the vehicles can always reset the origin of the position measurements.

\subsection{Optimal Control Problem}
MPC generates control signal sequences $\{\textbf{u}^{(i)}\}_{i \in \mathcal{M}} = \big\{\big(u^{(i)}(0), \dots, u^{(i)}(N-1)\big)\big\}_{i \in \mathcal{M}}$ by solving an optimization problem that minimizes a cost function over the prediction horizon $N$, subject to the constraints \eqref{eq:vel_constraints}-\eqref{eq:pos_constraints}, and the vehicle models \big(\eqref{eq:PWA_mod} or \eqref{eq:MLD_mod}\big).
The first elements of these control sequences $\big\{u^{(i)}(0)\big\}_{i \in \mathcal{M}}$ are then applied, and the optimization is performed again at the next time step in a receding horizon fashion.
Define the centralized MPC optimization problem as
\begin{equation} \label{eq:cent_MPC}
	\begin{aligned} 
		\min_{\{(\textbf{u}^{(i)}, \textbf{x}^{(i)}, \textbf{j}^{(i)})\}_{i \in \mathcal{M}}} &\sum_{k=0}^{N} \Big( \|x^{(l)}(k) - r(k)\|_{Q_x} \\
		 &\hspace{-0.8cm}+  \sum_{i \in \mathcal{M} \setminus \{1\}} \|x^{(i)}(k) - x^{(i-1)}(k) - \eta\big(x^{(i)}(k)\big)\|_{Q_x} \Big) \\
		&\hspace{-0.8cm}+ \sum_{i \in \mathcal{M}} \sum_{k=0}^{N-1} \| u^{(i)}(k) \|_{Q_u} \\
		&\hspace{-0.8cm}\text{s.t.} \quad \text{for} \: i\in \mathcal{M} \\
		&\hspace{-0.8cm}\quad \eqref{eq:vel_constraints}, \eqref{eq:safe_constraints}, \eqref{eq:pos_constraints}, \: k = 1,\dots,N \\ 
		&\hspace{-0.8cm}\quad  \text{vehicle model}, \eqref{eq:throttle_constraints}, \: k = 0,\dots,N-1 \\
		&\hspace{-0.8cm}\quad x^{(i)}(0) = x^{(i)}
	\end{aligned}
\end{equation}
where the bold $\textbf{u}/\textbf{x}/\textbf{j}$ represent the control, state, and gear decision variables over the prediction horizon, $Q_x$ and $Q_u$ are weight matrices, and $x^{(i)}$ is the current state of vehicle $i$.
The operator $\|\cdot\|$ is a general norm penalty, which will be specified in Section \ref{sec:experiments}.
\begin{remark}
	The gear decision variables $\textbf{j}$ enter \eqref{eq:cent_MPC} through the vehicle model, either explicitly as a discrete input in Model \RomanNumeralCaps{2}, or implicitly, dependent on the velocity, in Model \RomanNumeralCaps{1}.
	A possible augmentation of \eqref{eq:cent_MPC} would penalize gear changes explicitly in the cost to reduce fuel consumption \cite{liEcologicalAdaptiveCruise2020}.
\end{remark} 

The optimization problem \eqref{eq:cent_MPC} has coupling between the vehicles in the cost function and in the constraints.
For distributed MPC, the centralized MPC problem is broken into local subproblems for each vehicle $i \in \mathcal{M}$:
\begin{equation} \label{eq:local_MPC}
	\begin{aligned}
		\min_{\textbf{u}^{(i)}, \textbf{x}^{(i)}, \textbf{j}^{(i)}} &\sum_{k=0}^{N} \|x^{(i)}(k) - \bar{x}^{(i-1)}(k) - \eta\big(x^{(i)}(k)\big)\|_{Q_x} \\
		&+ \|\bar{x}^{(i+1)}(k) - x^{(i)}(k) -  \eta\big(\bar{x}^{(i+1)}(k)\big)\|_{Q_x} \\
			&+ \sum_{k=0}^{N-1} \| u^{(i)}(k) \|_{Q_u} \\
			&\text{s.t.} \: \eqref{eq:vel_constraints}, \eqref{eq:pos_constraints}, \: k = 1,\dots,N \\
			&\quad x_1^{(i)}(k) \leq \bar{x}_1^{(i - 1)}(k) - d_{\text{safe}}, \: k = 1,\dots,N \\
			&\quad \bar{x}_1^{(i + 1)}(k) \leq x_1^{(i)}(k) - d_{\text{safe}}, \: k = 1,\dots,N \\
			&\quad \text{vehicle model}, \eqref{eq:throttle_constraints}, \: k = 0,\dots,N-1 \\
			&\quad x^{(i)}(0) = x^{(i)}
	\end{aligned}
\end{equation}
where the new variables $\bar{\textbf{x}}^{(i-1)}$ and $\bar{\textbf{x}}^{(i+1)}$ represent some approximation or assumption on the trajectory of the vehicles in front of and behind vehicle $i$. 
The key consideration in distributed MPC approaches is then how vehicles decide or agree upon $\bar{\textbf{x}}^{(i-1)}$ and $\bar{\textbf{x}}^{(i+1)}$.

Naturally, the local subproblem for the leader uses the reference trajectory $r$ in the cost.
Likewise, the local subproblems for the front and rear vehicles will not include cost terms or safety constraints for a preceding or succeeding vehicle, respectively.
For brevity we do not present these subproblems explicitly.

\subsection{Feasibility}
\begin{color}{black}
We briefly discuss the feasibility of the optimization problems \eqref{eq:cent_MPC} and \eqref{eq:local_MPC}, considering each of the constraints introduced in Section \ref{sec:constraints}.
The position constraints \eqref{eq:pos_constraints} are only required to give a bounded state space; hence, $p_\text{MIN}$ and $p_\text{MAX}$ can be chosen arbitrarily small and large, respectively, such that the position constraints are always feasible.
The input constraints \eqref{eq:throttle_constraints} can always be satisfied as the control inputs $u^{(i)}(k)$ are independent decision variables.
The safe distance constraints \eqref{eq:safe_constraints} are coupled between adjacent vehicles, and are hence handled differently by different distributed controllers.
Consequently, these constraints are softened with slack variables that are penalized in the cost, such that the softened safe distance constraints can always be satisfied by making the slack variables non-zero.
The number of times that two vehicles breach the safe distance then becomes a comparison point between different distributed controllers.

The velocity and acceleration constraints \eqref{eq:vel_constraints} require closer attention.
For brevity we describe an $x_2(k)$ value such that $v_{\text{MIN}} \leq x_2(k) \leq v_{\text{MAX}}$ as a `viable' $x_2(k)$.
Note that for all viable $x_2(k)$, the existence of a control input $u(k)$ such that $x_2(k+1) = x_2(k)$ is sufficient to ensure the feasibility of both the velocity and acceleration constraints \eqref{eq:vel_constraints} for time steps after $k$.
We then provide a sufficient condition on the vehicle mass $m$ such that this control input exists.
For the PWA model, i.e., Model \RomanNumeralCaps{1}, the velocity dynamics are
\begin{equation}
	\begin{aligned}
		x_2(k+1) = &x_2(k) + T\Big( -\frac{1}{m} \hat{f}\big(x_2(k)\big) \\
		 &- \mu g + \hat{b}\big(x_2(k)\big) u(k)\Big).
	\end{aligned}
\end{equation}
Solving for $x_2(k+1) = x_2(k)$ gives
\begin{equation}\label{eq:u_const}
	u(k) = \big(\frac{1}{m}\hat{f}\big(x_2(k)\big) + \mu g\big)/\hat{b}\big(x_2(k)\big),
\end{equation}
with $u(k) > 0$ as all terms on the right-hand side are positive.
We must then check only that $u(k)$ in \eqref{eq:u_const} is less than $u_\text{MAX}$, i.e., for all viable $x_2$
\begin{equation}\label{eq:m_min_pwa}
	m \geq \frac{\hat{f}(x_2)}{\hat{b}(x_2) u_\text{MAX} - \mu g}.
\end{equation}
As $\hat{b}$ is constant within each PWA region and $\hat{f}$ is monotonically increasing with $x_2$ (see Figure \ref{fig:fric}), it is sufficient to evaluate the bound \eqref{eq:m_min_pwa} at the maximum $x_2$ value within each PWA region, taking the largest bound as the requirement on $m$.
Following the same steps for the discrete-input model, Model \RomanNumeralCaps{2}, it is sufficient to bound $m$ for every valid gear for a given $x_2$, i.e., for viable $x_2$, and for all $j$ such that $v_{j, \text{L}} \leq x_2 \leq v_{j, \text{H}}$,
\begin{equation}\label{eq:m_min_disc}
	m \geq \frac{\hat{f}(x_2)}{b(j, x_2) u_\text{MAX} - \mu g}.
\end{equation}
By the same arguments as above, it is sufficient to evaluate the bound \eqref{eq:m_min_disc} for the maximum $x_2$ of each gear, i.e., $x_2 = v_{j, \text{H}}$, and within each friction PWA region, i.e., $x_2 = \alpha$ and $x_2 = \dot{s}_\text{MAX}$, taking the largest bound for $m$.
For the vehicle parameters listed in this paper (see Section \ref{sec:experiments}), this gives $m \geq 2.82$ kg for both models, such that for vehicle masses $m \geq 2.82$ kg, and viable $x_2(k)$, the constraints \eqref{eq:vel_constraints} are feasible for time steps after $k$.
The final consideration to make is that, due to model errors between the prediction models and the real vehicle models, $x_2(0)$ could be outside the viable range, in which case a bound on the model error is required to determine the existence of a $u(0)$ such that $x_2(1)$ is viable.
Alternatively, using a bounded model error, robust MPC techniques \cite{chisciSystemsPersistentDisturbances2001a} such as constraint tightening could be applied to ensure $x_2(0)$ is always viable. 
Quantifying the model mismatch is beyond the scope of this paper, and we instead assume that the horizon is long enough and the sample time is small enough such that $x_2(0)$ remains in the viable range, such that the velocity and acceleration constraints \eqref{eq:vel_constraints} are feasible for all $k$.
\end{color}

\subsection{Performance Measures}
To compare the performance of different controllers on the benchmark problem, consider the tracking performance
\begin{equation} \label{eq:tracking}
	 \begin{aligned}
	J = &\sum_{k=0}^{K_\text{SIM}} \bigg( \|x^{(l)}(k) - r(k)\|_{Q_x} \\
	&+  \sum_{i \in \mathcal{M} \setminus \{1\}}  \|x^{(i)}(k) - x^{(i-1)}(k) - \eta\big(x^{(i)}(k)\big)\|_{Q_x}  \\
	&+  \sum_{i \in \mathcal{M}} \|u^{(i)}(k) \|_{Q_u} \bigg),
	\end{aligned}
\end{equation}
where $K_\text{SIM}$ is the length of a simulation.
This measure evaluates how well the vehicles track each other, through penalizing the state differences, and how efficiently, through penalizing the throttle input.
Additionally, the number of times the safe distance $d_\text{safe}$ between vehicles is breached can be considered, indicating when there is insufficient agreement between vehicles on the values of coupled states.
In Section \ref{sec:experiments} we also evaluate the properties of the controllers that influence their applicability: computation time, amount of communication between vehicles, and the local memory required for solving the optimization problems.

\begin{color}{black}
\subsection{Open Source Code}
An open source code base for the benchmark problem is available at \url{https://github.com/SamuelMallick/hybrid-vehicle-platoon.git} under the GNU General Public license.
The code base includes the simulation environment, the two vehicle models presented in this paper, and implementations of all controllers compared in this paper.
The current controllers use the Gurobi \cite {gurobi} optimizer to solve the MPC optimal control problems.
Alternative vehicle models and control strategies can be added in a plug-and-play manner.
\end{color}

\section{Hybrid MPC Controllers} \label{sec:controllers}
In this section we introduce five hybrid MPC controllers that are evaluated on the benchmark problem.
These are:
\begin{itemize}
	\item Centralized MPC
	\item Decentralized MPC
	\item Sequential distributed MPC
	\item Event-based distributed MPC
	\item ADMM-based distributed MPC.
\end{itemize}
For each controller, both Model \RomanNumeralCaps{1} and \RomanNumeralCaps{2} can be used as the prediction model.
When Model \RomanNumeralCaps{1} is used, it is converted into MLD form.
\begin{color}{black} Hence, all MPC optimization problems in the following are MILPs or MIQPs, and can be solved with a branch-and-bound (BnB) solver. 
BnB is a widely used methodology for solving combinatorial optimization problems, involving enumerating possible solutions to the problem and storing partial solutions in a tree structure.
The tree structure branches, creating subproblems, by partitioning the solution space, while bounds on the optimum can be used to prune entire parts of the tree, where the solutions are known to be suboptimal.
In mixed-integer optimization the subproblems are created by relaxing integer variables to be continuous.
Branching is then done by partitioning the continuous feasible region.
For an extended overview of BnB, in particular for mixed-integer programming, see \cite{morrisonBranchandBoundAlgorithmsSurvey2016}.
\end{color}

\subsection{Centralized}
The centralized MPC controller solves (\ref{eq:cent_MPC}) directly at each time step, finding all control inputs for all vehicles in a single optimization problem.
Using a BnB solver, the global optimum of (\ref{eq:cent_MPC}) can be found.
As such, this controller serves as the baseline performance for the following distributed controllers.

\subsection{Decentralized MPC}
The decentralized MPC controller finds control inputs by solving the local subproblems (\ref{eq:local_MPC}) in parallel, without any communication between the vehicles.
Decentralized MPC has been explored for linear systems \cite{venkatDistributedMPCStrategies2008}, and can trivially be extended to hybrid systems, with the only difference being the prediction model in the local MPC problems.
\begin{color}{black}While traditional decentralized MPC \textit{ignores} coupling, for the platoon control case we take advantage of a vehicle's ability to measure, in an error-free way, the state of adjacent vehicles.\end{color}
At each time step vehicles measure the position and velocity of adjacent vehicles and, assuming constant velocities\footnote{\begin{color}{black}We also implemented and tested simple vehicle speed prediction algorithms using linear function and saturation function predictors. We found that, in our simulations, the constant speed assumption outperformed both these. 
As the main focus of this work is the formulation and proposal of the benchmark problem, an exploration of vehicle speed prediction algorithms is left to future work, and only constant speed predictions are used for the decentralized controller.\end{color}}, extrapolate the positions to generate the estimates $\bar{\textbf{x}}^{(i-1)}$ and $\bar{\textbf{x}}^{(i+1)}$.

\subsection{Sequential distributed MPC}
Sequential distributed MPC approaches solve the local subproblems serially in a fixed sequence.
After a subsystem solves its subproblem, it communicates the solution to the next subsystems  in the sequence.
This idea has been implemented for linear systems in \cite{richardsRobustDistributedModel2007}, and hybrid systems in \cite{kuwataDistributedRobustReceding2007}.

For a platoon, the natural sequence of subproblems is the leader first, followed by the adjacent vehicles, i.e., vehicle $l$ first, then vehicles $l-1$ and $l+1$, then vehicles $l-2$ and $l+2$, and so on until all vehicles have solved their subproblems.
When vehicle $i$ solves its subproblem, it communicates its predicted state to vehicle $i-1$ and $i+1$.
Therefore, vehicles have perfect knowledge of $\bar{\textbf{x}}^{(i-1)}$ or $\bar{\textbf{x}}^{(i+1)}$, when solving (\ref{eq:local_MPC}), if vehicles $i-1$ or $i+1$ were earlier in the sequence than vehicle $i$.
For $\bar{\textbf{x}}^{(i-1)}$ or $\bar{\textbf{x}}^{(i+1)}$ of vehicles $i-1$ or $i+1$ after vehicle $i$ in the sequence, vehicle $i$ uses the solutions for $\bar{\textbf{x}}^{(i-1)}$ or $\bar{\textbf{x}}^{(i+1)}$ from the previous time step, shifted by one time step, assuming constant velocity for the shifting.

\subsection{Event-based distributed MPC}
Event-based distributed MPC \cite{grossDistributedPredictiveControl2013} involves each subsystem solving local problems in parallel, and communicating the solutions only in the event of a significant cost improvement.
The local problems are enlarged subproblems that considers the control inputs and states of coupled subsystems as decision variables.
\begin{color}{black}
In the platoon, coupled subsystems are adjacent vehicles, and each vehicle solves a subproblem considering the states and inputs of the preceding and succeeding vehicles as additional decision variables. \end{color}
Starting from some base solution, all vehicles solve their subproblems in parallel, and compute a cost-improvement from the base solution.
If a cost-improvement threshold is achieved, the vehicle with the highest cost improvement communicates its optimized trajectories to other vehicles, that then serve as the new base solutions, and the process is repeated.
In this way the method uses `parallel computation and serial communication' to achieve agreement and improvement on the shared variables.
The base solutions at each time step are the shifted solutions from the previous time step, again extrapolating positions assuming constant velocity.
For the initial time step of a simulation, when no previous solutions are available, vehicles use measurement to generate a base solution, as in the decentralized approach.

Originally, the approach would continue to iterate computation and event-based communication until the sampling time is exhausted \cite{grossDistributedPredictiveControl2013}.  
In this comparison, where the computation time is not restricted, we instead run the algorithm for a range of fixed numbers of iterations. 

As vehicles optimize over the trajectories of adjacent vehicles, this approach assumes that vehicles have exact knowledge of the dynamics of their adjacent vehicles. 
This may not always be a reasonable assumption, particularly in the case of heterogeneous platoons.

\subsection{ADMM-based distributed MPC}
\begin{table*}
	\centering
	\color{black}
	\caption{Experimental parameters}
	\label{tab:constraints}
	\begin{tabular}{ccccccccccccccc}
		\hline
		\textbf{Parameter} & $g$ & $\mu$ & $c$ & $v_{\text{MIN}}$ & 	$v_{\text{MAX}}$ & $a_{\text{MIN}}$ & $a_{\text{MAX}}$ & $u_{\text{MAX}}$ & $d_{\text{safe}}$ & $p_{\text{MIN}}$ & $p_{\text{MAX}}$ & $Q_x$ & $Q_u$ & $T$ \\
		 \textbf{Value} & $9.8$ & $0.01$ & $0.5$  & 3.94 & 45.84  & -2  & 2.5  & 1 & 25 & 0 & 10000 &$\begin{bmatrix}
		 	1 & 0 \\ 0 & 0.1
		 \end{bmatrix}$ & 1 & 1  \\
		 \hline
		 \textbf{Parameter} & \multicolumn{2}{c}{$\substack{\text{task 1} \\ (d_0, t_0, l)}$} & \multicolumn{2}{c}{$\substack{\text{task 2} \\ (d_0, t_0, l)}$} & \multicolumn{2}{c}{$\substack{\text{task 3} \\ (d_0, t_0, l)}$} & \multicolumn{2}{c}{$\substack{\text{task 1} \\ m_i}$} & \multicolumn{3}{c}{$\substack{\text{task 2} \\ m_i}$} & \multicolumn{3}{c}{$\substack{\text{task 3} \\ m_i}$} \\
		 \textbf{Value}  & \multicolumn{2}{c}{$(50, \text{NA}, 1)$}  & \multicolumn{2}{c}{$(10, 3, 1)$}  & \multicolumn{2}{c}{$(10, 3, \mathcal{M} \setminus \{1\})$}  & \multicolumn{2}{c}{$m_i = 800$}  & \multicolumn{3}{c}{$m_i \sim \mathcal{U}(700, 1000)$} & \multicolumn{3}{c}{$m_i \sim \mathcal{U}(700, 1000)$} \\
		 \hline
		 \textbf{Parameter} & \multicolumn{6}{c}{Initial positions} & \multicolumn{6}{c}{Initial velocities} \\
		 \textbf{Value} & \multicolumn{6}{c}{$x_1^{(i-1)}(0) - x_1^{(i)}(0) \sim \mathcal{U}(60, 160)$} & \multicolumn{6}{c}{$x_2^{(i)}(0) \sim \mathcal{U}(5, 35)$} \\
		\hline
	\end{tabular}
\end{table*}
While ADMM is only guaranteed to converge for convex optimization problems, in the literature it has been applied to hybrid, non-convex, control problems `naively', in the hope that, while not guaranteed to converge, the resulting solution will be close to the true optimum \cite{luanDecompositionDistributedOptimization2020}. An ADMM approach involves augmenting the local cost function in (\ref{eq:local_MPC}) with penalty terms that penalize the difference between the assumed coupled variables $\bar{\textbf{x}}^{(i-1)}$ and $\bar{\textbf{x}}^{(i+1)}$ and their true values.
Vehicles iteratively solve their subproblems and communicate the solutions to adjacent vehicles.
In the convex case, at convergence, the coupled variables will be agreed upon across the network, and the local optimizers will converge to the optimizers of the centralized optimization problem.

Stopping criteria for the ADMM iterations can be constructed from convergence of the coupled variables to common values \cite{boydDistributedOptimizationStatistical2010}.
However, in the hybrid case, where convergence is not guaranteed, we will use a range of fixed numbers of iterations.
The reader is referred to \cite{boydDistributedOptimizationStatistical2010} for a detailed overview of ADMM, to \cite{summersDistributedModelPredictive2012} for an example of distributed MPC via ADMM for linear systems, and to \cite{luanDecompositionDistributedOptimization2020} for an example of ADMM being applied to distributed hybrid MPC.

\section{Experiments} \label{sec:experiments}
\begin{color}{black}
\begin{figure*}
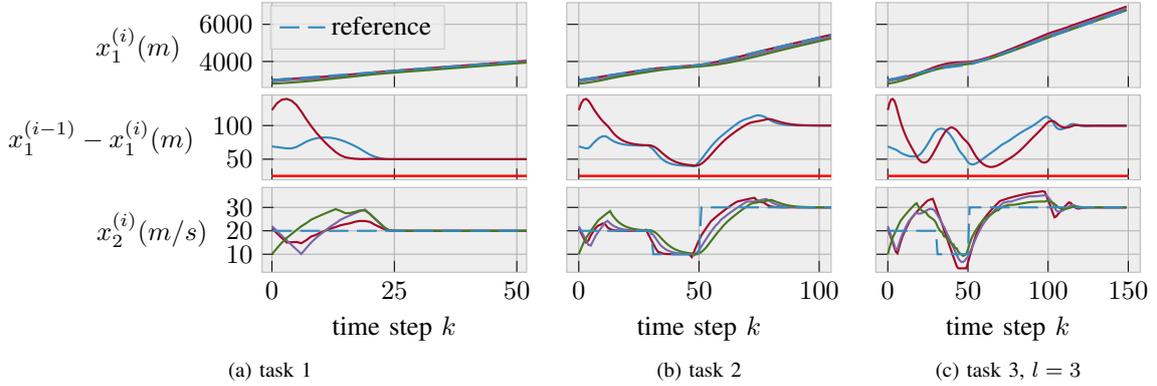

	\centering
	\color{black}
	\subfloat[task 1 \label{fig:task_1}]{\input{media/tikz/tasks/task_1}}
	\subfloat[task 2 \label{fig:task_2}]{\input{media/tikz/tasks/task_2}}
	\subfloat[task 3, $l = 3$ \label{fig:task_3}]{\input{media/tikz/tasks/task_3}}
	\caption{Platoon formation under the centralized controller for the 3 tasks with $M = 3$ and $N = 6$. , Position (top), inter-vehicle spacing (middle) with safe distance (red line), and velocity (bottom).}
	\label{fig:tasks}
\end{figure*}
In this section we compare the five controllers presented in Section \ref{sec:controllers} on three instances the benchmark problem.
The first and simplest instance, referred to as task 1, considers the front vehicle as the leader, a constant speed reference trajectory, homogenous vehicles, and a constant distance spacing policy.
With task 1 we explore the performance and complexity of the controllers under the different prediction models (Model \RomanNumeralCaps{1} and \RomanNumeralCaps{2}) and norm choices $\|\cdot\|$ in the costs.
The second instance, task 2, considers an aggressive variable-speed reference trajectory, inhomogeneous vehicles, a velocity-dependent spacing policy, and again the front vehicle as leader.
Finally, the third instance, task 3, is identical to task 2, except that the controllers are evaluated with  each vehicle as the leader, excluding the front vehicle.
On tasks 2 and 3 we explore the performance and characteristics of the controllers as the size of the platoon ($M$) and the length of the prediction horizon ($N$) vary.
All tasks consider each vehicle to be initialized with a randomized velocity, distributed uniformly in the range $\begin{bmatrix}
	5 & 35
\end{bmatrix} \text{ms}^{-1}$, and randomized positions behind the preceding vehicle, distributed uniformly in the range $\begin{bmatrix}
60 & 160
\end{bmatrix} \text{m}$ .

Table \ref{tab:constraints} gives all numerical coefficients that are used in the experiments for the tasks, controllers, and constraints.
Figure \ref{fig:tasks} demonstrates representative platoon trajectories for the three tasks.
We highlight that these are just three of many possible configurations for the benchmark problem, with the tuning knobs, outlined in Section \ref{sec:benchmark}, providing a way to construct many more instances of varying difficulty.

We consider the following performance indicators:
\begin{itemize}
	\item Tracking performance $J$ \eqref{eq:tracking}.
	\item Tracking performance with respect to the centralized controller $\Delta J = J - J_\text{cent}$. This indicates the performance loss introduced by a distributed controller, with respect to the performance of the centralized controller.
	\item Computation time $t_\text{COMP} = (t_\text{min}, t_\text{av}, t_\text{max})$. This triple contains the minimum, average, and maximum time required to compute the control inputs for a simulation step.
	These times are for the \textit{entire platoon}, i.e., for the decentralized controller, the computation time for a time step is the maximum computation time required to compute a local MPC solution, as the subsystems solve their subproblems in parallel, while for the sequential controller, it is the sum of the local computation times in the sequential order.
	\item Number of explored nodes $n_\text{no}$. This indicator serves as a proxy for the local memory required by each controller. It is determined by the maximum number of nodes that are explored in one BnB procedure.
\end{itemize}
In all simulations, the continuous-time nonlinear hybrid model \eqref{eq:ODE} is used to simulate the underlying system.
All mixed-integer programs are solved using Gurobi \cite{gurobi}.
All simulations are run on a Linux machine using five AMD EPYC 7252 cores, 1.38GHz clock speed, and 251Gb of RAM.
In the following, for the iterative controllers, the number of iterations is presented along with the controller name, e.g., ADMM(20) represents the ADMM-based controller with 20 iterations.
\end{color}

\subsection{Linear versus Quadratic Costs}

\begin{figure*}
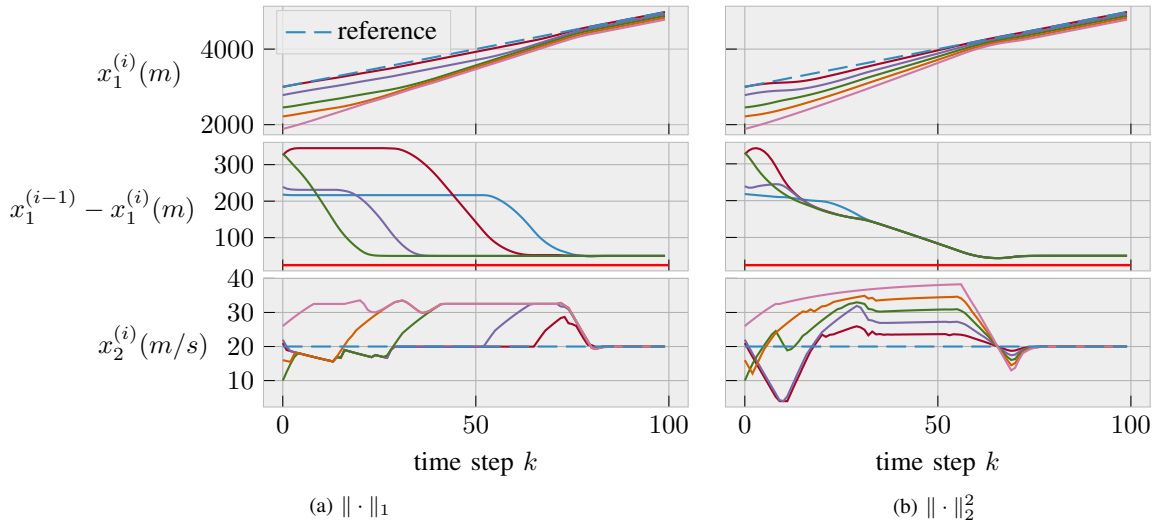

	\centering
	\subfloat[$\|\cdot\|_1$ \label{fig:1_norm}]{\input{media/tikz/norm_comp/1_norm}}
	\subfloat[$\|\cdot\|_2^2$ \label{fig:2_norm}]{\input{media/tikz/norm_comp/2_norm}}
	\caption{Platoon formation under the centralized controller with 1- and 2-norm costs for $M = 5$ and $N = 6$.
	Position (top), inter-vehicle spacing (middle) with safe distance (red line), and velocity (bottom).}
	\label{fig:norms}
\end{figure*}

First, we compare the use of 1- and 2-norm costs, $\|\cdot\| = \|\cdot\|_1$ and $\|\cdot\| = \|\cdot\|_2^2$, respectively.
Figure \ref{fig:norms} compares the platoon trajectory on task 1, with a centralized controller, for each norm.
Quadratic, 2-norm-based, costs penalize large tracking errors heavily, and small tracking errors lightly, while 1-norm-based costs penalize the tracking errors linearly.
As a result, the platoon formation differs significantly between the two.
A 2-norm-based cost encourages the fleet to cooperate and form the platoon earlier, while moving towards the reference trajectory.

\begin{figure}
	\centering
	\input{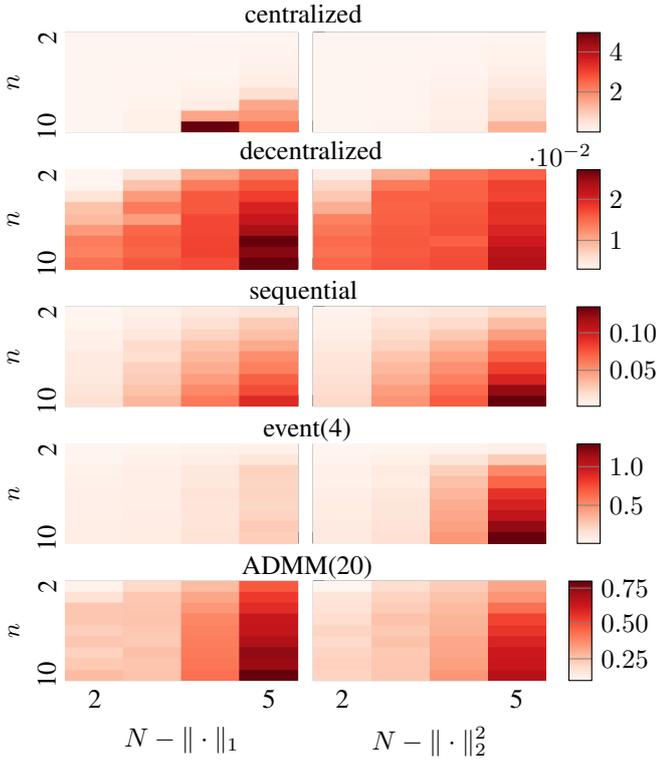}
	\caption{Heat map for the average computation time of each controller, in seconds, under 1- and 2-norm stage costs. The length of the prediction horizon increases along the x-axis, while the size of the network increases down the y-axis.}
	\label{fig:MILP_MIQP}
\end{figure}

Figure \ref{fig:MILP_MIQP} shows a heat-map comparison of the average computation time $t_\text{av}$ for each controller under each norm, as the prediction horizon and size of the network are increased.
The heat map shows that for the sequential and decentralized controllers, the 1-norm-based cost provides a computational benefit, while for the other controllers, it is the 2-norm-based cost that provides the speed up.
It is often cited that, when solving mixed-integer programs, 1-norm-based costs are preferred, as MILPs can be solved faster than MIQPs by commercial BnB solvers \cite{coronaAdaptiveCruiseControl2008}.
We highlight, however, that the time required to solve a mixed-integer program varies significantly from problem to problem, based on the initial conditions, warm-start solution, and the effectiveness of the online branching and pruning procedures.
As the different norms lead to different trajectories, and therefore different mixed-integer optimization problems, the choice should be explored explicitly for different controllers and control tasks.
In the remaining experiments in this paper, we use costs based on the 2-norm, as the faster convergence to a regularly spaced platoon is preferred.

\subsection{Model Comparison}

We compare the performance and computational complexity of the prediction models; Model \RomanNumeralCaps{1} and Model \RomanNumeralCaps{2}, using task 1.
Additionally, we compare the hybrid models against the use of a discretization of the nonlinear hybrid dynamics \eqref{eq:ODE} as a prediction model\footnote{Gurobi supports quadratic equality constraints and is thus able to solve the mixed-integer nonlinear programs (MINLP) that arise when using \eqref{eq:ODE} as a prediction model. We also compared Gurobi against the dedicated MINLP solver MindtPy \cite{bernal2018mixed} for solving the control problems, and found Gurobi to be faster.}.

Table \ref{tab:gear_mod_perf} gives the percentage difference between the models' tracking performances and average computation times on task 1 with $M=3$ and $N=5$.
For all controllers, the tracking performance is improved by using Model \RomanNumeralCaps{2} over Model \RomanNumeralCaps{1}.
This is in line with the observation that Model \RomanNumeralCaps{2}, by modeling the gear choice as a discrete input, retains the use of a larger subset of the velocity-gear decision space.
As Model \RomanNumeralCaps{2} is not restricted to have a one-to-one mapping between velocity and gear, for a given velocity there may be a gear choice available to Model \RomanNumeralCaps{2} that is not available to Model \RomanNumeralCaps{1}, which may improve tracking by, e.g., giving a faster acceleration.
Hence, for a given state, any optimal solution to the MPC optimization problem using Model \RomanNumeralCaps{1} provides an upper bound on the cost of the optimal solution when using Model \RomanNumeralCaps{2}.

The average computation time required for Model \RomanNumeralCaps{2} is significantly higher than that for Model \RomanNumeralCaps{1}.
This is particularly severe for the control strategies that use MPC controllers with decision spaces covering more than one vehicle, i.e., the centralized and event-based controllers. 
This large increase in computational complexity indicates that Model \RomanNumeralCaps{1}, the PWA model, contains some structure that is advantageous in the BnB process.
Indeed, for the centralized controller, the optimal control problem at each time step contains 105 and 120 binary decision variables for Model \RomanNumeralCaps{1} and Model \RomanNumeralCaps{2}, respectively.
However, following the Gurobi pre-solve procedure\footnote{The Gurobi pre-solve procedure attempts to simplify an optimization problem, prior to solving it. This process involves many complex operations, such as the removal of redundant constraints and variables, detection of implied bounds on variables, cut generation, etc.}, on average the numbers of remaining binary decision variables are 20 and 52 for Model \RomanNumeralCaps{1} and Model \RomanNumeralCaps{2}, respectively.
While slightly decreasing performance via a restriction of the usable velocity-gear space, Model \RomanNumeralCaps{1} offers significant computational advantages through its PWA structure.
An additional advantage of Model \RomanNumeralCaps{1} is that it can be converted to many equivalent modeling frameworks \cite{heemelsEquivalenceHybridDynamical2001a}, for which alternative control strategies could be explored.
These alternative strategies are beyond the scope of this paper and are left to future work.
In contrast, using Model \RomanNumeralCaps{2} as a prediction model will always result in a mixed-integer program, as the gears are explicitly modeled as discrete inputs.
In the following experiments, we use Model \RomanNumeralCaps{1} for all prediction models due to the significantly lower computational burden.

\begin{color}{black}The nonlinear prediction model gives a slight improvement in performance over Model \RomanNumeralCaps{2}, as it captures the true quadratic friction.
However, the addition of the non-convex constraints in the dynamics results in a vastly increased complexity.
Again this is particularly extreme for the centralized and event-based controllers, which solve larger optimization problems.
This motivates the use of a hybrid approximation for the nonlinear friction, sacrificing very little performance for a huge computational speed-up.

\begin{table*}
	\color{black}
	\centering
	\begin{tabular}{|c|c|c|c|c|c|}
		\hline
		& Centralized & Decentralized & Sequential & Event (4) & ADMM (20) \\
		\hline
		$\frac{J_\text{II} - J_\text{I}}{J_\text{I}}\times 100$ & $-4.06$  & $-3.77$ & $-4.03$ & $-4.05$ & $-3.64$ \\
		\hline
		$\frac{t_\text{av, II} - t_\text{av, I}}{t_\text{I}}\times 100$ & $1.58\mathrm{e}4$  & $243.20$ & $247.28$ & $1.42\mathrm{e}4$ & $159.42$ \\
		\hline
		\hline
		$\frac{J_\text{NL} - J_\text{I}}{J_\text{I}}\times 100$ & $-4.56$  & $-4.14$ & $-4.54$ & $-4.26$ & $-4.06$ \\
		\hline
		$\frac{t_\text{av, NL} - t_\text{av, I}}{t_\text{I}}\times 100$ & $5.32\mathrm{e}5$  & $1.36\mathrm{e}3$ & $1.63\mathrm{e}3$ & $6.96\mathrm{e}4$ & $813.00$ \\
		\hline
	\end{tabular}
	\caption{Comparison between prediction models for task 1 with $M = 3$, $N = 5$.}
	\label{tab:gear_mod_perf}
\end{table*}

\subsection{Performance Comparison - Task 2}

We compare the performance of the controllers on task 2 as the size of the platoon $M$ and the prediction horizon $N$ vary.
For the platoon size, task 2 is simulated for a fixed prediction horizon of $N=6$, with the number of vehicles varying from $M = 2$ to $M = 10$.
Figure \ref{fig:n_sweep} presents the performance indicators over ten randomized initial conditions and vehicle masses (see Table \ref{tab:constraints}).
The relative tracking performance $\Delta J$ shows that, in general, the non-centralized controllers introduce a performance drop with respect to centralized control and that, as the number of vehicles increases, the extent of this performance drop worsens. 
The event-based controllers perform the best for small numbers of vehicles, achieving centralized performance.
As the number of vehicles increases a performance drop is seen.
However, this performance drop is not uniformly increasing with the number of vehicles, and is not uniformly improved by the number of iterations.
Particularly noticeable, the event(4) controller performs the best on average for $N = 7$ to $N = 9$, and the worst for $N = 10$.
This demonstrates the heuristic nature of this approach, as the performance drop is inconsistent, and more iterations of the algorithm do not necessarily result in a better performance.
In contrast, the other non-centralized controllers introduce a performance drop for all platoon sizes, that worsens as the number of vehicles increases.
The decentralized controller introduces extreme performance drops for large numbers of vehicles, demonstrating the importance of communication.

The computation times $t_\text{COMP}$ show that in general, as the number of vehicles increases, the computational complexity of the controllers increases.
Although the non-centralized controllers solve subproblems whose number of decision variables is independent of the platoon size, the computational requirements increase with the size of the platoon.
This demonstrates how the complexity of solving mixed-integer programs can vary with the complexity of the control challenge. 
Only the decentralized, sequential, and ADMM(5) controllers maintained a maximum computation time for the fleet $t_\text{max}$ of less than the sampling time $T = 1$s for all platoon sizes.
The event-based controllers in particular show extreme computational complexities, reaching $t_\text{max}$ greater than 100s. 
Note that the computation times for the event-based controllers are similar across numbers of iterations.
This shows that the extra iterations are used infrequently, as the extra iterations are only used only in the event of a cost improvement (see Section \ref{sec:controllers}).

The node count $n_\text{no}$ shows that the local memory required by the controllers is independent of the platoon size for the decentralized, sequential, and ADMM controllers.
For the event-based controllers, the amount of local memory required is significant, with respect to the other controllers, as the local mixed-integer problems consider the decision variables of adjacent vehicles, and the memory requirements scale poorly with the number of vehicles.

\begin{figure}
	\centering
	\hspace{1cm}\subfloat{
\begin{tikzpicture}

\newcommand{\myW}{0.6*\axisdefaultwidth}
\newcommand{\myH}{0.8*\axisdefaultheight}

\definecolor{chocolate213940}{RGB}{213,94,0}
\definecolor{cornflowerblue86180233}{RGB}{86,180,233}
\definecolor{darkcyan0158115}{RGB}{0,158,115}
\definecolor{darkgray178}{RGB}{178,178,178}
\definecolor{darkolivegreen7012033}{RGB}{70,120,33}
\definecolor{firebrick166640}{RGB}{166,6,40}
\definecolor{lightgray204}{RGB}{204,204,204}
\definecolor{palevioletred204121167}{RGB}{204,121,167}
\definecolor{silver188}{RGB}{188,188,188}
\definecolor{slategray122104166}{RGB}{122,104,166}
\definecolor{steelblue52138189}{RGB}{52,138,189}
\definecolor{whitesmoke238}{RGB}{238,238,238}

\begin{axis}[
width=\myW,
height=\myH,
axis background/.style={fill=whitesmoke238},
axis line style={silver188},
hide x axis,
hide y axis,
legend cell align={left},
legend columns=3,
legend style={
  fill opacity=0.8,
  draw opacity=1,
  text opacity=1,
  at={(0.5,0.5)},
  anchor=center,
  draw=lightgray204,
  fill=whitesmoke238
},
tick pos=left,
x grid style={darkgray178},
xmajorgrids,
xmin=100, xmax=101,
xtick style={color=black},
xtick={4.6,4.8,5,5.2,5.4},
xticklabels={
  \(\displaystyle {4.6}\),
  \(\displaystyle {4.8}\),
  \(\displaystyle {5.0}\),
  \(\displaystyle {5.2}\),
  \(\displaystyle {5.4}\)
},
y grid style={darkgray178},
ymajorgrids,
ymin=4.725, ymax=5.275,
ytick style={color=black},
ytick={4.5,4.75,5,5.25,5.5},
yticklabels={
  \(\displaystyle {4.50}\),
  \(\displaystyle {4.75}\),
  \(\displaystyle {5.00}\),
  \(\displaystyle {5.25}\),
  \(\displaystyle {5.50}\)
}
]
\addplot [thick, steelblue52138189, mark=o, mark size=3, mark options={solid,fill opacity=0}]
table {%
5 5
};
\addlegendentry{decentralized}
\addplot [thick, firebrick166640, mark=x, mark size=3, mark options={solid,fill opacity=0}]
table {%
5 5
};
\addlegendentry{sequential}
\addplot [thick, slategray122104166, dash pattern=on 2pt off 3.3pt, mark=triangle, mark size=3, mark options={solid,rotate=180,fill opacity=0}]
table {%
5 5
};
\addlegendentry{event(4)}
\addplot [thick, darkolivegreen7012033, dash pattern=on 2pt off 3.3pt, mark=triangle, mark size=3, mark options={solid,rotate=180,fill opacity=0}]
table {%
5 5
};
\addlegendentry{event(6)}
\addplot [thick, chocolate213940, dash pattern=on 2pt off 3.3pt, mark=triangle, mark size=3, mark options={solid,rotate=180,fill opacity=0}]
table {%
5 5
};
\addlegendentry{event(10)}
\addplot [thick, palevioletred204121167, dash pattern=on 7.4pt off 3.2pt, mark=square, mark size=3, mark options={solid,fill opacity=0}]
table {%
5 5
};
\addlegendentry{ADMM(5)}
\addplot [thick, cornflowerblue86180233, dash pattern=on 7.4pt off 3.2pt, mark=square, mark size=3, mark options={solid,fill opacity=0}]
table {%
5 5
};
\addlegendentry{ADMM(20)}
\addplot [thick, darkcyan0158115, dash pattern=on 7.4pt off 3.2pt, mark=square, mark size=3, mark options={solid,fill opacity=0}]
table {%
5 5
};
\addlegendentry{ADMM(50)}

\end{axis}
\end{tikzpicture}}
	\vspace{-0.3cm}
	\subfloat{\input{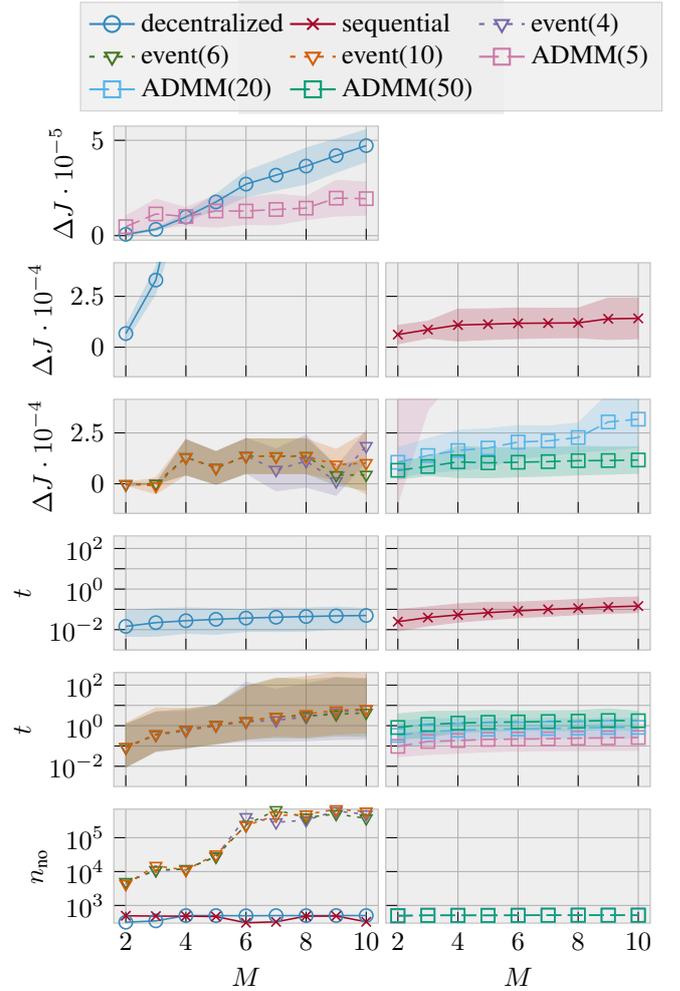}}
	\caption{Task 2 with prediction horizon $N = 6$. Rows 1-3: average relative closed-loop tracking performance $\Delta J$ (a standard deviation in shaded region). Rows 2 and 3 share a y-axis scaling such that controllers in different plots can be compared. As the decentralized and ADMM(5) controllers have vastly increased performance drops, Row 1 shows a zoomed out view for these.
	Rows 4-5: average computation time $t_\text{av}$ ($t_\text{max}$ and $t_\text{min}$ in shaded region) - log scale.
	Row 6: node count $n_\text{no}$ - log scale.}
	\label{fig:n_sweep}
\end{figure}

For the varying prediction horizon, task 2 is simulated for a fixed platoon size of $M = 6$, with the prediction horizon varying from $N=2$ to $N=6$.
Figure \ref{fig:N_sweep} presents the performance indicators over ten randomized initial conditions and vehicle masses (see Table \ref{tab:constraints}).
The tracking performance shows that as the size of the prediction horizon increases, in general, the performance of the non-centralized controllers improves.
A clear exception is the decentralized controller, which performs worse as the horizon increases.
Exploring this further, Figure \ref{fig:N_sweep_decent} shows the decentralized performance as $N$ increases for $M = 2$.
The opposite trend is observed, indicating a non-trivial relationship between the horizon length and platoon size for the decentralized controller.
This may result from the decentralized controller relying heavily on extrapolated trajectory estimates, such that for some platoon sizes, a shorter prediction horizon becomes beneficial, reducing prediction uncertainty.
The computation times $t_\text{COMP}$ show exponential increases in computational complexity as the prediction horizon increases (linear trend on a log scale).
This demonstrates how the worst-case complexity of a mixed-integer program grows exponentially with the number of integer decision variables \cite{bemporadControlSystemsIntegrating1999}, and that as the prediction horizon increases, the subproblems have increasing numbers of binary decision variables.
Similarly, the node counts $n_\text{no}$ shows that the local memory requirements also grow exponentially as the prediction horizon increases.
These effects are most significant in the event-based controllers, where the subproblems consider the decision variables associated with more than one vehicle.
The computation time and memory requirements for the event-based controllers are nearly identical, indicating that extra iterations were rarely used.

All controllers satisfied the safety distance constraints for all task 2 simulations.

\begin{figure}
	\centering
	\hspace{1cm}\subfloat{
\begin{tikzpicture}

\newcommand{\myW}{0.6*\axisdefaultwidth}
\newcommand{\myH}{0.8*\axisdefaultheight}

\definecolor{chocolate213940}{RGB}{213,94,0}
\definecolor{cornflowerblue86180233}{RGB}{86,180,233}
\definecolor{darkcyan0158115}{RGB}{0,158,115}
\definecolor{darkgray178}{RGB}{178,178,178}
\definecolor{darkolivegreen7012033}{RGB}{70,120,33}
\definecolor{firebrick166640}{RGB}{166,6,40}
\definecolor{lightgray204}{RGB}{204,204,204}
\definecolor{palevioletred204121167}{RGB}{204,121,167}
\definecolor{silver188}{RGB}{188,188,188}
\definecolor{slategray122104166}{RGB}{122,104,166}
\definecolor{steelblue52138189}{RGB}{52,138,189}
\definecolor{whitesmoke238}{RGB}{238,238,238}

\begin{axis}[
width=\myW,
height=\myH,
axis background/.style={fill=whitesmoke238},
axis line style={silver188},
hide x axis,
hide y axis,
legend cell align={left},
legend columns=3,
legend style={
  fill opacity=0.8,
  draw opacity=1,
  text opacity=1,
  at={(0.5,0.5)},
  anchor=center,
  draw=lightgray204,
  fill=whitesmoke238
},
tick pos=left,
x grid style={darkgray178},
xmajorgrids,
xmin=100, xmax=101,
xtick style={color=black},
xtick={4.6,4.8,5,5.2,5.4},
xticklabels={
  \(\displaystyle {4.6}\),
  \(\displaystyle {4.8}\),
  \(\displaystyle {5.0}\),
  \(\displaystyle {5.2}\),
  \(\displaystyle {5.4}\)
},
y grid style={darkgray178},
ymajorgrids,
ymin=4.725, ymax=5.275,
ytick style={color=black},
ytick={4.5,4.75,5,5.25,5.5},
yticklabels={
  \(\displaystyle {4.50}\),
  \(\displaystyle {4.75}\),
  \(\displaystyle {5.00}\),
  \(\displaystyle {5.25}\),
  \(\displaystyle {5.50}\)
}
]
\addplot [thick, steelblue52138189, mark=o, mark size=3, mark options={solid,fill opacity=0}]
table {%
5 5
};
\addlegendentry{decentralized}
\addplot [thick, firebrick166640, mark=x, mark size=3, mark options={solid,fill opacity=0}]
table {%
5 5
};
\addlegendentry{sequential}
\addplot [thick, slategray122104166, dash pattern=on 2pt off 3.3pt, mark=triangle, mark size=3, mark options={solid,rotate=180,fill opacity=0}]
table {%
5 5
};
\addlegendentry{event(4)}
\addplot [thick, darkolivegreen7012033, dash pattern=on 2pt off 3.3pt, mark=triangle, mark size=3, mark options={solid,rotate=180,fill opacity=0}]
table {%
5 5
};
\addlegendentry{event(6)}
\addplot [thick, chocolate213940, dash pattern=on 2pt off 3.3pt, mark=triangle, mark size=3, mark options={solid,rotate=180,fill opacity=0}]
table {%
5 5
};
\addlegendentry{event(10)}
\addplot [thick, palevioletred204121167, dash pattern=on 7.4pt off 3.2pt, mark=square, mark size=3, mark options={solid,fill opacity=0}]
table {%
5 5
};
\addlegendentry{ADMM(5)}
\addplot [thick, cornflowerblue86180233, dash pattern=on 7.4pt off 3.2pt, mark=square, mark size=3, mark options={solid,fill opacity=0}]
table {%
5 5
};
\addlegendentry{ADMM(20)}
\addplot [thick, darkcyan0158115, dash pattern=on 7.4pt off 3.2pt, mark=square, mark size=3, mark options={solid,fill opacity=0}]
table {%
5 5
};
\addlegendentry{ADMM(50)}

\end{axis}
\end{tikzpicture}}
	\vspace{-0.3cm}
	\subfloat{\input{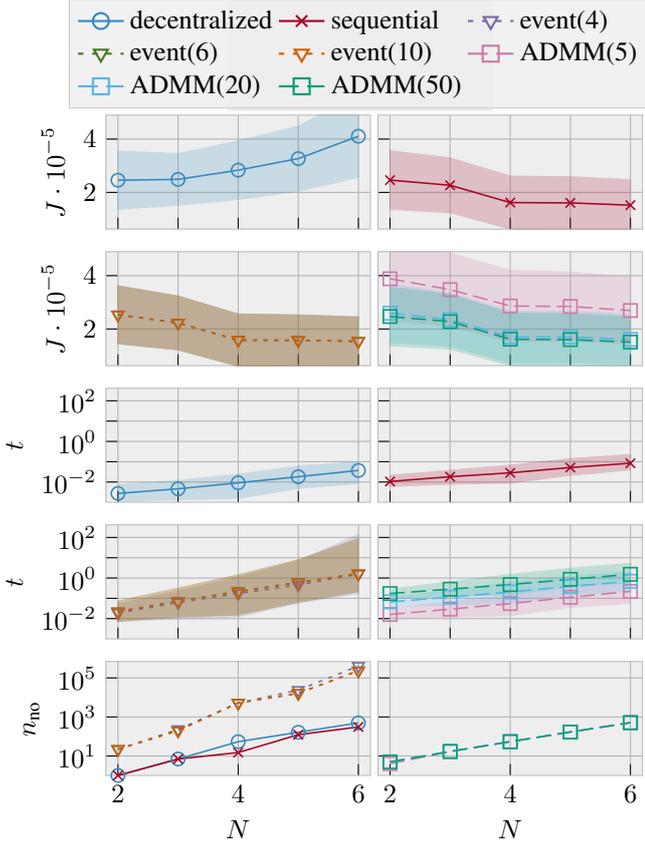}}
	\caption{{Task 2 with number of vehicles $M = 6$. Rows 1-2: average closed-loop tracking performance $J$ (a standard deviation in shaded region).
	Rows 3-4: average computation time $t_\text{av}$ ($t_\text{max}$ and $t_\text{min}$ in shaded region) - log scale.
	Row 5: node count $n_\text{no}$ - log scale.}}
	\label{fig:N_sweep}
\end{figure}
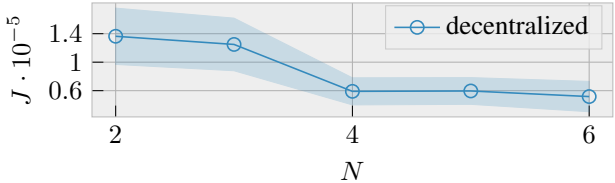
\begin{figure}
	\centering
\begin{tikzpicture}
	
\newcommand{\myW}{1*\axisdefaultwidth}
\newcommand{\myH}{0.8*\axisdefaultheight}

\definecolor{darkgray178}{RGB}{178,178,178}
\definecolor{lightgray204}{RGB}{204,204,204}
\definecolor{silver188}{RGB}{188,188,188}
\definecolor{steelblue52138189}{RGB}{52,138,189}
\definecolor{whitesmoke238}{RGB}{238,238,238}

\begin{axis}[
width=\myW,
height=\myH,
axis background/.style={fill=whitesmoke238},
axis line style={silver188},
legend cell align={left},
legend style={fill opacity=0.8, draw opacity=1, text opacity=1, draw=lightgray204, fill=whitesmoke238},
tick pos=left,
scaled ticks=false,
x grid style={darkgray178},
xlabel={\(\displaystyle N\)},
xmajorgrids,
xmin=1.8, xmax=6.2,
xtick style={color=black},
xtick={2,4,6},
xticklabels={
  \(\displaystyle {2}\),
  \(\displaystyle {4}\),
  \(\displaystyle {6}\),
},
y grid style={darkgray178},
ylabel={\(\displaystyle J \cdot 10^{-5}\)},
ymajorgrids,
ymin=23139.7257520798, ymax=183182.689746184,
ytick style={color=black},
ytick={20000,60000,100000,140000},
yticklabels={
  \(\displaystyle {0.2}\),
  \(\displaystyle {0.6}\),
  \(\displaystyle {1}\),
  \(\displaystyle {1.4}\),
}
]
\path [draw=steelblue52138189, fill=steelblue52138189, opacity=0.2, very thin]
(axis cs:2,175908.009564634)
--(axis cs:2,96833.7641902446)
--(axis cs:3,87956.8605150048)
--(axis cs:4,40002.9962069837)
--(axis cs:5,40793.4431220623)
--(axis cs:6,30414.40593363)
--(axis cs:6,73018.5172469699)
--(axis cs:6,73018.5172469699)
--(axis cs:5,78353.6637598776)
--(axis cs:4,78075.1462193886)
--(axis cs:3,161851.227929047)
--(axis cs:2,175908.009564634)
--cycle;

\addplot [semithick, steelblue52138189, mark=o, mark size=2.5, mark options={solid,fill opacity=0}]
table {%
2 136370.886877439
3 124904.044222026
4 59039.0712131861
5 59573.55344097
6 51716.4615903
};
\addlegendentry{decentralized}
\end{axis}

\end{tikzpicture}
	\caption{Average closed-loop tracking performance $J$ (a standard deviation in shaded region) for the decentralized controller on task 2 with $M = 2$.}
	\label{fig:N_sweep_decent}
\end{figure}

\subsection{Performance Comparison - Task 3}
We compare the performance of a subset of the controllers on task 3 as the size of the platoon $M$ increases.
Task 3 is simulated with a prediction horizon of $N = 6$ and the number of vehicles varying from $M = 2$ to $M = 5$.
For each value of $M$, a simulation is run where each vehicle, except the front vehicle, is the leader, i.e., $l \in \mathcal{M} \setminus \{1\}$.
For task 3, where all vehicles can be the leader, we note that the sequential controller is often unable to track the reference trajectory, due to the strict agreement enforced on coupled states. 
In the sequential approach, solutions for the coupled variables are passed along the sequence of vehicles, with no opportunity given for negotiation.
As such, with the leader `sandwiched' in the middle of the platoon, it may be unable to coerce the platoon towards the reference trajectory, as the other vehicles, focusing on tracking adjacent vehicles, may restrict the leader.
Figure \ref{fig:seq_task_3} demonstrates an example of this. While the leader $l=4$ tries to speed up to reach the reference trajectory, pushing up against its preceding vehicle, the vehicles at the front of the platoon slow down to track each other.
By the velocity-dependent spacing policy, the desired spacing shrinks as the vehicles slow down; causing them to slow down more and trapping the leader, up to the point where a deadlock is reached and the leader cannot initiate any acceleration, as it is at the safe distance limit with its adjacent vehicles.
\begin{figure}
	\centering
	\input{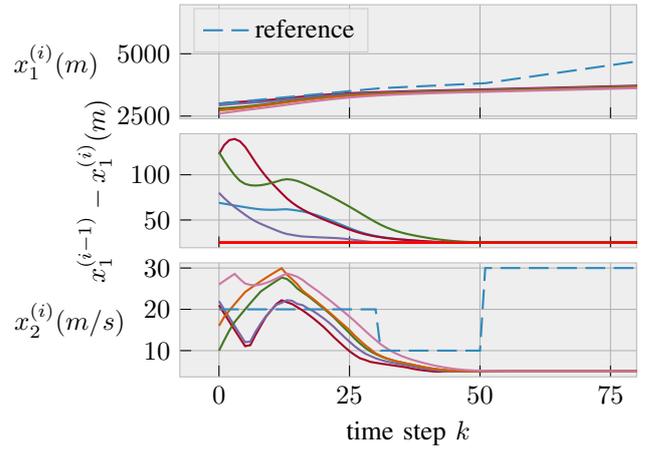}
	\caption{Sequential controller for task 3 with $M = 5$, $N = 6$, and $l = 4$.
	Position (top), inter-vehicle spacing (middle) with safe distance (red line), and velocity (bottom).}
	\label{fig:seq_task_3}
\end{figure}

Figure \ref{fig:multi_lead} presents the performance indicators averaged over all leader choices.
The relative tracking performance shows that as the number of vehicles increases the performance drop of the non-centralized controllers increases.
Furthermore, the performance drop is more severe than for task 2, demonstrating the added difficulty of task 3.
In particular, on this more complex task, the ADMM(50) controller is outperformed by the decentralized controller for $M \geq 4$, demonstrating how ADMM in the non-convex setting gives no guarantee of approaching the centralized optimum, with the suboptimality worsening with task complexity.
The event-based controller performs the best; however, the performance drop scales poorly with the number of vehicles, and the complexity of the controller can be seen to explode even for modest numbers of vehicles, with $t_\text{max} > 100$ even for $M = 5$.
The inter-vehicle safe distance was maintained successfully for all task 3 simulations.

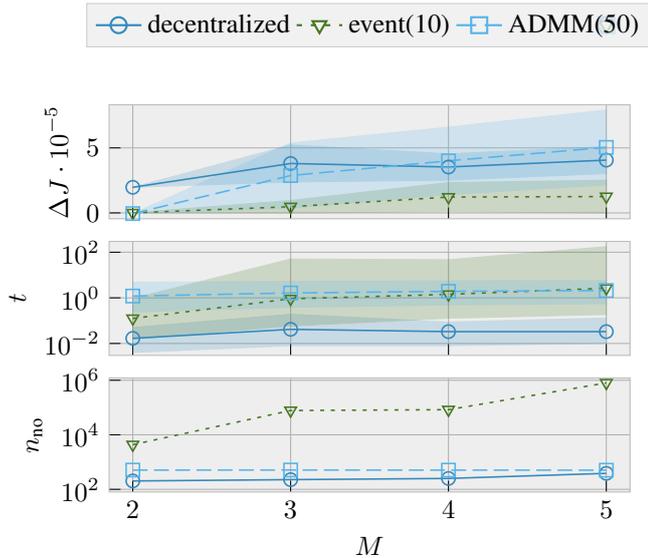
\begin{figure}
	\centering
\begin{tikzpicture}
	
\newcommand{\myW}{\axisdefaultwidth}
\newcommand{\myH}{0.8*\axisdefaultheight}

\definecolor{cornflowerblue86180233}{RGB}{86,180,233}
\definecolor{darkgray178}{RGB}{178,178,178}
\definecolor{darkolivegreen7012033}{RGB}{70,120,33}
\definecolor{lightgray204}{RGB}{204,204,204}
\definecolor{silver188}{RGB}{188,188,188}
\definecolor{steelblue52138189}{RGB}{52,138,189}
\definecolor{whitesmoke238}{RGB}{238,238,238}

\begin{groupplot}[group style={group size=1 by 4, vertical sep = 0.3cm}]
\nextgroupplot[
width=\myW,
height=\myH,
axis background/.style={fill opacity=0, draw opacity=0},
axis line style={silver188},
hide x axis,
hide y axis,
legend cell align={left},
legend columns=3,
legend style={
  fill opacity=0.8,
  draw opacity=1,
  text opacity=1,
  at={(0.5,0.5)},
  anchor=center,
  draw=lightgray204,
  fill=whitesmoke238
},
tick pos=left,
tick scale binop=\times,
scaled ticks = false,
x grid style={darkgray178},
xmajorgrids,
xmin=1.85, xmax=5.15,
xtick style={color=black},
xticklabels={},
y grid style={darkgray178},
ymajorgrids,
ymin=4.725, ymax=5.275,
ytick style={color=black},
ytick={4.5,4.75,5,5.25,5.5},
yticklabels={
  \(\displaystyle {4.50}\),
  \(\displaystyle {4.75}\),
  \(\displaystyle {5.00}\),
  \(\displaystyle {5.25}\),
  \(\displaystyle {5.50}\)
}
]
\addplot [thick, steelblue52138189, mark=o, mark size=3, mark options={solid,fill opacity=0}]
table {%
5 5
};
\addlegendentry{decentralized}
\addplot [thick, darkolivegreen7012033, dash pattern=on 2pt off 3.3pt, mark=triangle, mark size=3, mark options={solid,rotate=180,fill opacity=0}]
table {%
5 5
};
\addlegendentry{event(10)}
\addplot [thick, cornflowerblue86180233, dash pattern=on 7.4pt off 3.2pt, mark=square, mark size=3, mark options={solid,fill opacity=0}]
table {%
5 5
};
\addlegendentry{ADMM(50)}

\nextgroupplot[
width=\myW,
height=\myH,
axis background/.style={fill=whitesmoke238},
axis line style={silver188},
scaled x ticks=manual:{}{\pgfmathparse{#1}},
tick pos=left,
scaled ticks = false,
x grid style={darkgray178},
xmajorgrids,
xmin=1.85, xmax=5.15,
xtick style={color=black},
xtick={2,3,4,5},
xticklabels={},
y grid style={darkgray178},
ylabel={\(\displaystyle \Delta J \cdot 10^{-5}\)},
ymajorgrids,
ymin=-45136.8507176487, ymax=831124.911430553,
ytick style={color=black},
ytick={-500000,0,500000,1000000},
yticklabels={
  \(\displaystyle {\ensuremath{-}500000}\),
  \(\displaystyle {0}\),
  \(\displaystyle {5}\),
  \(\displaystyle {10}\)
}
]
\path [draw=steelblue52138189, fill=steelblue52138189, opacity=0.2, very thin]
(axis cs:2,196793.731869311)
--(axis cs:2,196793.731869311)
--(axis cs:3,237748.404142358)
--(axis cs:4,251643.217293871)
--(axis cs:5,301796.874796452)
--(axis cs:5,510066.484015254)
--(axis cs:5,510066.484015254)
--(axis cs:4,454104.434488499)
--(axis cs:3,522205.764711808)
--(axis cs:2,196793.731869311)
--cycle;

\path [draw=darkolivegreen7012033, fill=darkolivegreen7012033, opacity=0.2, very thin]
(axis cs:2,653.174389154141)
--(axis cs:2,653.174389154141)
--(axis cs:3,2.73015810389188)
--(axis cs:4,10157.8688934777)
--(axis cs:5,330.13613879522)
--(axis cs:5,251194.815497911)
--(axis cs:5,251194.815497911)
--(axis cs:4,233159.386161052)
--(axis cs:3,95607.7755447971)
--(axis cs:2,653.174389154141)
--cycle;

\path [draw=cornflowerblue86180233, fill=cornflowerblue86180233, opacity=0.2, very thin]
(axis cs:2,-5306.7706200032)
--(axis cs:2,-5306.7706200032)
--(axis cs:3,36734.0498223238)
--(axis cs:4,140956.229919141)
--(axis cs:5,216284.54161828)
--(axis cs:5,791294.831332908)
--(axis cs:5,791294.831332908)
--(axis cs:4,659582.689243445)
--(axis cs:3,537557.705715713)
--(axis cs:2,-5306.7706200032)
--cycle;

\addplot [semithick, steelblue52138189, mark=o, mark size=2.5, mark options={solid,fill opacity=0}]
table {%
2 196793.731869311
3 379977.084427083
4 352873.825891185
5 405931.679405853
};
\addplot [semithick, darkolivegreen7012033, dash pattern=on 1.5pt off 2.475pt, mark=triangle, mark size=2.5, mark options={solid,rotate=180,fill opacity=0}]
table {%
2 653.174389154141
3 47805.2528514505
4 121658.627527265
5 125762.475818353
};
\addplot [semithick, cornflowerblue86180233, dash pattern=on 5.55pt off 2.4pt, mark=square, mark size=2.5, mark options={solid,fill opacity=0}]
table {%
2 -5306.7706200032
3 287145.877769019
4 400269.459581293
5 503789.686475594
};

\nextgroupplot[
width=\myW,
height=\myH,
axis background/.style={fill=whitesmoke238},
axis line style={silver188},
log basis y={10},
scaled x ticks=manual:{}{\pgfmathparse{#1}},
tick pos=left,
tick scale binop=\times,
x grid style={darkgray178},
xmajorgrids,
xmin=1.85, xmax=5.15,
xtick style={color=black},
xtick={2,3,4,5},
xticklabels={},
y grid style={darkgray178},
ylabel={\(\displaystyle t\)},
ymajorgrids,
ymin=0.003, ymax=298.012425070785,
ymode=log,
ytick style={color=black},
ytick={0.01, 1, 100},
yticklabels={
  \(\displaystyle {10^{-2}}\),
  \(\displaystyle {10^{0}}\),
  \(\displaystyle {10^2}\),
}
]
\path [draw=steelblue52138189, fill=steelblue52138189, opacity=0.2, very thin]
(axis cs:2,0.0499999523162842)
--(axis cs:2,0.00399994850158691)
--(axis cs:3,0.00799989700317383)
--(axis cs:4,0.00800013542175293)
--(axis cs:5,0.00999999046325684)
--(axis cs:5,0.13100004196167)
--(axis cs:5,0.13100004196167)
--(axis cs:4,0.0869998931884766)
--(axis cs:3,0.191999912261963)
--(axis cs:2,0.0499999523162842)
--cycle;

\path [draw=darkolivegreen7012033, fill=darkolivegreen7012033, opacity=0.2, very thin]
(axis cs:2,0.927922964096069)
--(axis cs:2,0.0154440402984619)
--(axis cs:3,0.0588340759277344)
--(axis cs:4,0.127741813659668)
--(axis cs:5,0.184901237487793)
--(axis cs:5,174.673008918762)
--(axis cs:5,174.673008918762)
--(axis cs:4,46.9857950210571)
--(axis cs:3,50.5066320896149)
--(axis cs:2,0.927922964096069)
--cycle;

\path [draw=cornflowerblue86180233, fill=cornflowerblue86180233, opacity=0.2, very thin]
(axis cs:2,4.8447105884552)
--(axis cs:2,0.239450693130493)
--(axis cs:3,0.384573221206665)
--(axis cs:4,0.487899541854858)
--(axis cs:5,0.554888248443604)
--(axis cs:5,5.70071506500244)
--(axis cs:5,5.70071506500244)
--(axis cs:4,5.58235836029053)
--(axis cs:3,5.23272800445557)
--(axis cs:2,4.8447105884552)
--cycle;

\addplot [semithick, steelblue52138189, mark=o, mark size=2.5, mark options={solid,fill opacity=0}]
table {%
2 0.0167533318201701
3 0.0416100056966146
4 0.0330066691504584
5 0.0330016656716665
};
\addplot [semithick, darkolivegreen7012033, dash pattern=on 1.5pt off 2.475pt, mark=triangle, mark size=2.5, mark options={solid,rotate=180,fill opacity=0}]
table {%
2 0.124700415929159
3 0.910429216225942
4 1.41956619739532
5 2.65901135047277
};
\addplot [semithick, cornflowerblue86180233, dash pattern=on 5.55pt off 2.4pt, mark=square, mark size=2.5, mark options={solid,fill opacity=0}]
table {%
2 1.19903385003408
3 1.63463017702103
4 1.92718291812473
5 2.09713585535685
};

\nextgroupplot[
width=\myW,
height=\myH,
axis background/.style={fill=whitesmoke238},
axis line style={silver188},
log basis y={10},
tick pos=left,
tick scale binop=\times,
x grid style={darkgray178},
xmajorgrids,
xmin=1.85, xmax=5.15,
xtick style={color=black},
xtick={2,3,4,5},
xticklabels={
  \(\displaystyle {2}\),
  \(\displaystyle {3}\),
  \(\displaystyle {4}\),
  \(\displaystyle {5}\),
},
xlabel=\(M\),
y grid style={darkgray178},
ylabel={\(\displaystyle n_\text{no}\)},
ymajorgrids,
ymin=80, ymax=1207757.09760808,
ymode=log,
ytick style={color=black},
ytick={1,100,10000,1000000,100000000,10000000000},
yticklabels={
  \(\displaystyle {10^{0}}\),
  \(\displaystyle {10^{2}}\),
  \(\displaystyle {10^{4}}\),
  \(\displaystyle {10^{6}}\),
  \(\displaystyle {10^{8}}\),
  \(\displaystyle {10^{10}}\)
}
]
\addplot [semithick, steelblue52138189, mark=o, mark size=2.5, mark options={solid,fill opacity=0}]
table {%
2 203
3 227
4 251
5 386
};
\addplot [semithick, darkolivegreen7012033, dash pattern=on 1.5pt off 2.475pt, mark=triangle, mark size=2.5, mark options={solid,rotate=180,fill opacity=0}]
table {%
2 4344
3 77847
4 83293
5 798440
};
\addplot [semithick, cornflowerblue86180233, dash pattern=on 5.55pt off 2.4pt, mark=square, mark size=2.5, mark options={solid,fill opacity=0}]
table {%
2 509
3 512
4 508
5 505
};
\end{groupplot}

\end{tikzpicture}
	\caption{Task 3 with number of vehicles $M = 6$. Row 1: average relative closed-loop tracking performance $J$ (a standard deviation in shaded region).
	Row 2: average computation time $t_\text{av}$ ($t_\text{max}$ and $t_\text{min}$ in shaded region) - log scale.
	Row 3: node count $n_\text{no}$ - log scale.}
	\label{fig:multi_lead}
\end{figure}

\subsection{Comparison Summary}
In this section we provide a summary of the controller comparison, giving general guidelines of when each controller is preferred.
The decentralized controller presents the best option from the perspective of complexity, consistently requiring the least computation time and memory, thanks to subproblems being solved just once in parallel.
Furthermore, the decentralized controller requires no communication, thus removing the overhead of inter-vehicle communication infrastructure.
The cost for this low complexity is lower tracking performance, with the distributed controller introducing a performance drop that scales poorly with the size of the platoon. 
However, when computational power is limited, or communication is unavailable, the decentralized controller is preferred.

The sequential controller presents a slightly higher complexity than the decentralized controller, as the subproblems are solved in series rather than in parallel. 
Showcasing the benefit of even limited communication, the performance drop of the sequential controller is far smaller than that of the decentralized controller, particularly for large platoons.
However, the sequential controller's rigid agreement mechanism renders it inapplicable when the leader is not the front vehicle.
Therefore, when communication is available and the leader is the front vehicle, the sequential controller is recommended, giving a modest performance drop with a modest computational burden.

The event-based controllers overall give the best tracking performance.
However, the computational complexity is severe, with extremely large computation times and memory requirements, scaling poorly with the size of the platoon.
Furthermore, the performance does not necessarily improve with increasing numbers of iterations, potentially requiring manual tuning as platoon sizes and tasks vary. 
Additionally, the event-based controllers optimize over the trajectories of adjacent vehicles, requiring full knowledge of their dynamics.
Finally, communicating the updated solutions between iterations requires communication between vehicles up to two hops away.
Consequently, the event-based controllers are recommended when good tracking performance is desired at the expense of restrictive requirements on communication infrastructure, complexity, and information sharing.

The ADMM controllers present a complexity that increases with the number of algorithm iterations, with high numbers of iterations needed to achieve modest drops in tracking performance.
The iterative mechanism allows for negotiation, and the ADMM controllers can be applied when the leader is not the front vehicle.
As such, the ADMM controller is preferred over the sequential controller only when the leader is not the front vehicle. 
\end{color}

\section{Conclusions and Concluding Remarks} \label{sec:conclussion}
In this work we have presented a case study to serve as a benchmark problem for comparing distributed MPC controllers for hybrid systems. 
We have considered control of a platoon of vehicles with explicit gear management.
Two hybrid vehicle modeling options have been presented, the benchmark problem has been specified, and five existing hybrid MPC approaches have been evaluated on the benchmark.

\begin{color}{black}The performance of the evaluated existing controllers highlights the need for new distributed hybrid MPC strategies that can approach centralized performance, even as the size of the system increases.\end{color}
In particular, distributed controllers that strike a favorable trade-off between computational intensity and performance are clearly lacking.
Indeed, one point of note is that all controllers evaluated require the online solution to a mixed-integer linear or quadratic optimization problem.
This limits the applicability of these approaches as the scale of the problem increases.
We highlight that there is huge potential for the development of distributed hybrid MPC controllers that avoid solving mixed-integer programs online altogether.

The benchmark problem presented in this work can be modified through the tuning knobs: platoon size, reference trajectory, spacing policy, homogeneity, and leader position, for a modular and adaptable control problem, upon which future distributed hybrid MPC controllers can be evaluated. 
Future work will, in addition to developing controllers that fill the gaps outlined above, involve exploring distributed hybrid MPC strategies for this problem that can provide theoretical guarantees on recursive feasibility and stability, as well as string stability of the network.


\bibliographystyle{IEEEtran}
\bibliography{bib.bib}

\vspace*{-1cm}

\begin{IEEEbiography}[{\includegraphics[width=1in,height=1.25in,clip,keepaspectratio]{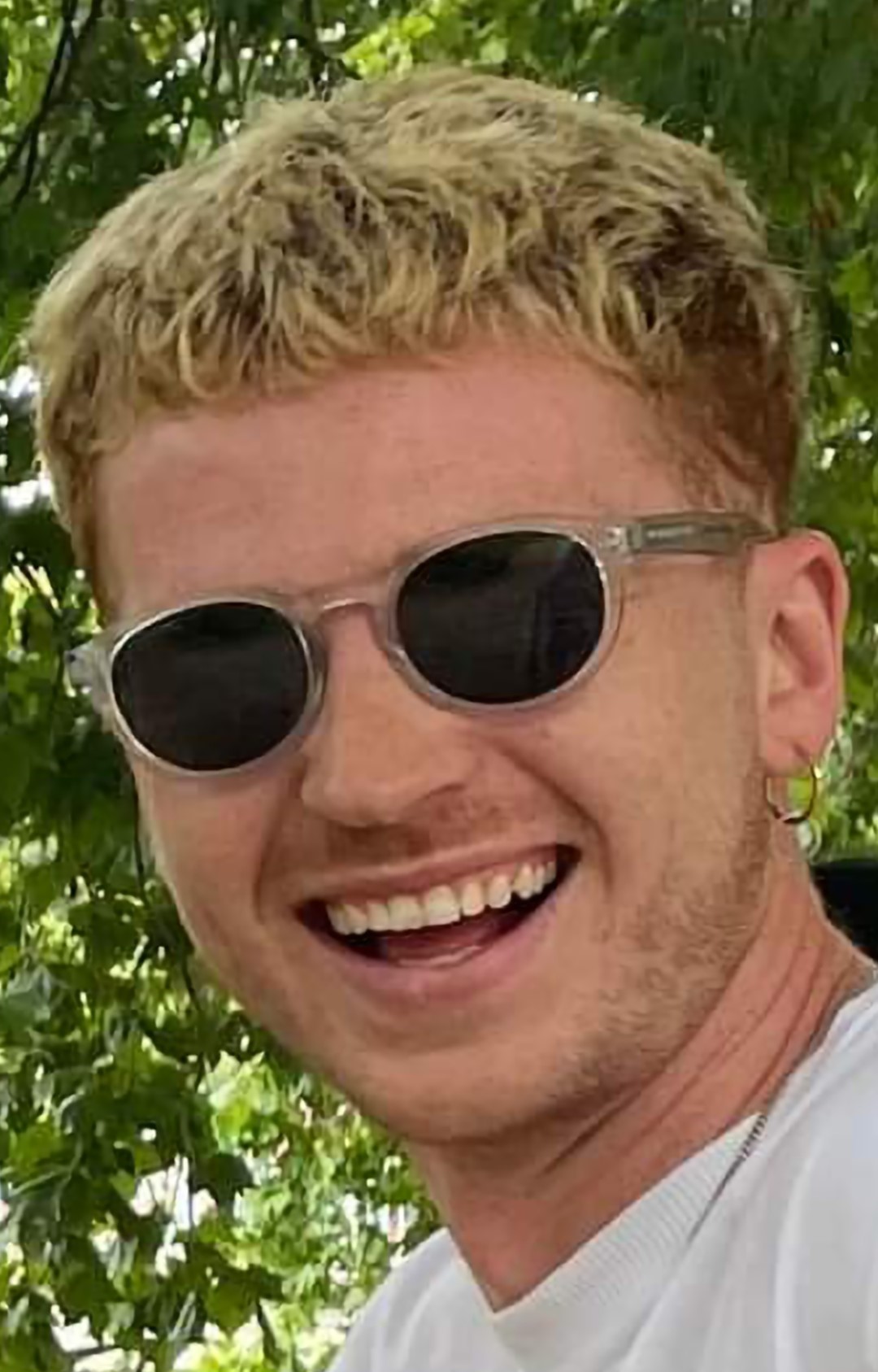}}]{Samuel Mallick} 
	received the B.Sc. and M.Sc. degrees from The University of Melbourne in 2020 and 2022, respectively. He is currently a Ph.D. candidate at the Delft Center for Systems and Control, Delft University of Technology, The Netherlands.
	
	His research interests include model predictive control, reinforcement learning, and distributed control of large-scale and hybrid systems.
\end{IEEEbiography}

\vspace*{-1cm}

\begin{IEEEbiography}[{\includegraphics[width=1in,height=1.25in,clip,keepaspectratio]{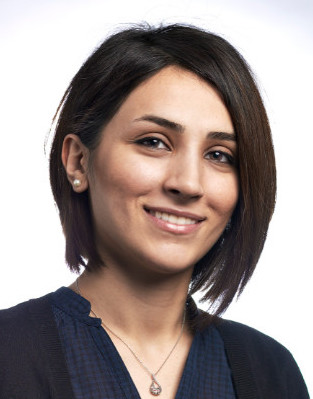}}]{Azita Dabiri} 
	received the Ph.D. degree from the Automatic Control Group, Chalmers University of Technology, in 2016. She was a Post-Doctoral Researcher with the Department of Transport and Planning, TU Delft, from 2017 to 2019. In 2019, she received an ERCIM Fellowship and also a Marie Curie Individual Fellowship, which allowed her to perform research at the Norwegian University of Technology (NTNU), as a Post-Doctoral Researcher, from 2019 to 2020, before joining the Delft Center for Systems and Control, TU Delft, as an Assistant Professor. Her research interests are in the areas of integration of model-based and learning-based control and its applications in transportation networks.
\end{IEEEbiography}
\vspace*{-1cm}

\begin{IEEEbiography}[{\includegraphics[width=1in,height=1.25in,clip,keepaspectratio]{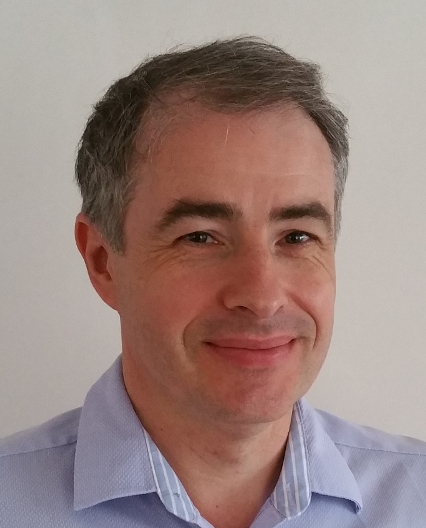}}]{Bart De Schutter}(Fellow, IEEE) 
	received the PhD degree (\emph{summa cum laude}) in applied sciences from KU Leuven, Belgium, in 1996. He is currently a Full Professor and Head of Department at the Delft Center for Systems and Control, Delft
	University of Technology, The Netherlands. His research interests include multi-level
	and multi-agent control, model predictive control, learning-based control, and control
	of hybrid systems, with applications in intelligent transportation systems and smart energy systems. 
	
	Prof.\ De Schutter is a Senior Editor of the IEEE Transactions on Intelligent Transportation Systems and an Associate Editor of the IEEE Transactions
	on Automatic Control.
\end{IEEEbiography}
\vfill

\end{document}